\documentclass[pra,twocolumn,superscriptaddress,amsmath,amssymb,aps,floatfix]{revtex4-2}
\usepackage{graphicx}
\usepackage{dcolumn}
\usepackage{bm}
\usepackage{amsmath,amssymb}
\usepackage{graphics}
\usepackage{amsfonts}
\usepackage{epsfig}
\usepackage{float}
\usepackage{hyperref}
\usepackage{adjustbox}
\usepackage{rotating}
\usepackage{float}
\usepackage{blindtext}
\usepackage{multirow}
\usepackage{graphicx}
\usepackage{dcolumn}
\usepackage{bm}
\usepackage{amsmath}
\usepackage{import}
\usepackage{physics}
\usepackage{verbatim}
\usepackage{listings}
\usepackage{xcolor}
\usepackage{xr-hyper}
\usepackage{hyperref}
\usepackage{cleveref}
\usepackage{newfloat}
\usepackage{caption}
\usepackage{subcaption}
\DeclareFloatingEnvironment[name={Fig S},fileext=lof]{suppfigure}
\begin{document}
\title{Recurrent generation of maximally entangled single particle states via quantum walks on cyclic graphs}
\author{Dinesh Kumar Panda}
\email{dineshkumar.quantum@gmail.com}
\author{Colin Benjamin}
\email{colin.nano@gmail.com}
\affiliation{School of Physical Sciences, National Institute of Science Education and Research Bhubaneswar, Jatni 752050, India}
\affiliation{Homi Bhabha National Institute, Training School Complex, Anushaktinagar, Mumbai
400094, India}
\begin{abstract}
Maximally entangled single-particle states (MESPS) are opening new possibilities in quantum technology as they have the potential to encode more information and are robust to decoherence compared to their nonlocal two-particle counterparts. We find that a single coin can generate MESPS at recurrent time steps (periodically) via discrete-time quantum walks on both $4$ and $8$ site cyclic graphs. This scheme is resource-saving with possibly the most straightforward experimental realization since the same coin is applied at each time step. We also show that recurrent MESPS can be generated on any arbitrary $k$ site cyclic graph, $k\in\{3,4,5,8\}$ via effective-single (Identity and arbitrary coin) or two coin evolution sequences. Beyond their use in fundamental research, we propose an application of the generated MESPS in quantum cryptography protocols. MESPS as cryptographic keys can strengthen quantum-secure communication.
\end{abstract}
\maketitle
\textit{Introduction.---} Hybrid or single particle entanglement (SPE) refers to the entanglement between different degrees of freedom such as spatial mode, polarization, and orbital angular momentum belonging to the same particle~\cite{aqt}. The quantum signature of SPE is contextuality which rules out realistic non-contextual hidden-variable theories and violates Bell's inequality. Though SPE lacks in non-locality, it has its significance plus advantages over the nonlocal or multi-particle entanglement~\cite{aqt,kochen1}. SPE enables encoding more information at the single particle level, is more robust against decoherence, and has simpler experimental implementation than its nonlocal bipartite counterpart~\cite{gratsea_lewenstein_dauphin_2020, aqt, fang}. SPE has significant applications in photonic quantum information processing and analysis of states of photons and elementary particles~\cite{aqt}. Since an entangled state can be transmitted employing a single particle, SPE is a proven resource to improve existing QKD (quantum key distribution) protocols like the BB84 for secret key sharing and a QKD scheme with single-particle entangled photons, see \cite{aqt, bb84}. Quantum joining, a physical process that allows the transfer of intra-particle entanglement between photons into a single output photon's hybrid entanglement and its inverse, has been reported, and it has applications in quantum networking~\cite{vit}. Photonic SPE states are potentially advantageous in optical quantum networks because they enable a more flexible network with every photon transmitted via a suitable channel~\cite{zhu_xiao_huo_xue_2020}. SPE has also been used in experimental tests of non-contextual hidden variable theories~\cite{aqt}.

A quantum walker (or particle) is represented by a wave function and obeys the quantum superposition principle, and this makes quantum walks (QWs) superior compared to their classical counterparts~\cite{aharonov_davidovich_zagury_1993}. A discrete-time quantum walk (DTQW) evolves by repeatedly applying two quantum operators: coin and shift. A quantum walk can be described on a 1D or 2D lattice and analogously on a cyclic graph with $k$ sites ($k$-cycle). For some detailed studies on QWs on $k$-cycles, see Refs.~\cite{cb-13, cb-14,expt-cyclicQW}. Ref.~\cite{expt-cyclicQW} reports on the experimental implementation of QW on cyclic graphs with photons using linear optical elements. A recent work~\cite{cb-ap} shows that it is possible to design an ordered or periodic QW by combining two chaotic QWs on $3-$ or $4-$cycle via Parrondo strategy~\cite{p1}. Intriguingly, the emergence of order from chaos and its inverse in QWs has applications in quantum cryptography~\cite{cb-ap}, quantum secure direct communication protocol~\cite{sspanda-qsdc}, and in developing theory of quantum chaos control~\cite{cb-22}.

Several manuscripts recently reported that DTQWs on 1D lines could be efficient tools to generate entangled single-particle states (SPS) or SPE, see Refs.~\cite{gratsea_lewenstein_dauphin_2020,li_yan_he_wang_2018,ch2012disorder,vieira_amorim_rigolin_2013,vieira_amorim_rigolin_2014,gratsea2020universal, fang, me-cb}. Refs.~\cite{fang,li_yan_he_wang_2018} report on the experimental realization of SPE generation. Ref.~\cite{me-cb} shows that by incorporating Parrondo sequences of coin-operators in 1D DTQWs, one can obtain phase-independent SPE and, in a particular case, maximal SPE independent of the initial state parameters for time steps of 3 and 5.

There has been no attempt to generate maximally entangled SPS (MESPS) and for that matter SPE in cyclic graphs. Also, seeing the versatility of DTQWs and the preeminent applicability of SPE, exploring different methods to generate highly or maximally entangled SPS via DTQWs is an important task, as it would contribute to extending the horizons of quantum technologies~\cite{aqt}. Our main aim in this work is to study the propensity of DTQWs on cyclic graphs in generating MESPS using a single coin. In addition, we also study MESPS generation using an effective-single coin (i.e., coin operator and Identity operator) or two coins in a deterministic evolution-operator sequence and their relation to ordered QW dynamics.

\begin{figure}[H]
\includegraphics[width=8.8cm,height=2cm]{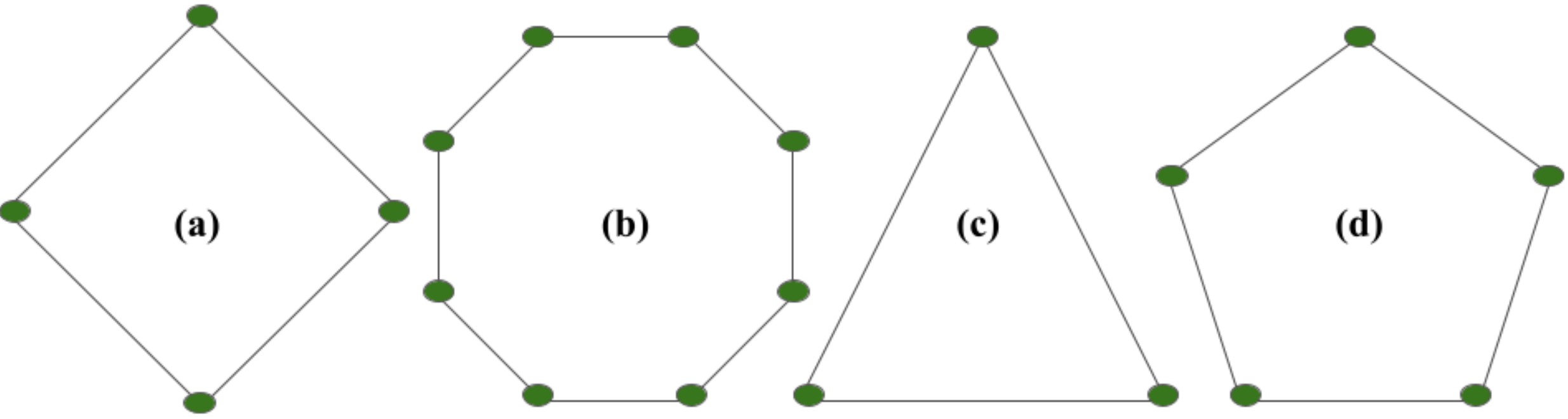}
\caption{4-cycle (a) and 8-cycle (b); 3-cycle (c) and 5-cycle (d), with sites marked by green dots.}
\label{f1}
\end{figure}

\textit{DTQW on cyclic graphs.---} A DTQW on a $k$-cycle (Fig.~\ref{f1}), is defined on a tensor product space ($H$) of position ($H_P$) and coin ($H_C$) Hilbert spaces, i.e., $H = H_P \otimes H_C $. $H_C$ is defined on the computational basis $\{\ket{0_c},\ket{1_c}\}$, whereas $H_P$ has the computational basis $\{\ket{x_p}: x_p \in \{0,1,2,...,k-1\}\}$. If the quantum walker is initially localized at the site $\ket{0_p}$ in a general superposition of the coin states, it is represented by $\ket{\psi_i}$ or $\ket{\psi(t=0)}$, i.e.,
\begin{equation}
\ket{\psi(t=0)} = \cos(\frac{\theta}{2})\ket{0_p,0_c} + e^{i\phi}\sin(\frac{\theta}{2})\ket{0_p,1_c},
\label{equ1}
\end{equation}
with $\theta \in [0, \pi]$ and $\phi \in [0, 2\pi)$.
The unitary coin operator is,
\begin{equation}
\hat{C}_{2}(\rho, \gamma, \eta) =
\begin{pmatrix}
\sqrt{\rho} & \sqrt{1-\rho}e^{i\gamma}\\
\sqrt{1-\rho}e^{i\eta} & -\sqrt{\rho}e^{i(\gamma+\eta)}
\end{pmatrix},
\label{equ2}
\end{equation}
\noindent
where $0\leq\rho\leq1$ \text{and} $0\leq\gamma,\eta\leq\pi$ .

The walker moves to the left (right) by one site for coin state $\ket{0_c}$ ($\ket{1_c}$). For the walker on $k$-cycle, we use the shift operator $\hat{S} = \sum_{q=0}^{1}\sum_{j=0}^{k-1}\ket{((j+2q-1) \text{ mod } k)_p}\bra{j_p}\otimes\ket{q_c}\bra{q_c}.$ The full evolution can be expressed as,
\begin{equation}
U_{k}(t) = \hat{S}.[I_k\otimes \hat{C}_{2}(\rho(t), \gamma(t), \eta(t))]\;,
\label{equ3}
\end{equation}
\noindent
where $I_k$ is a $k\cross k$ identity matrix.
The time-evolution of the system (quantum walker) after $t$ time steps is then,
\begin{eqnarray}
\label{equ4}
\ket{\psi(t)} & =& U_{k}(t)\ket{\psi(t-1)} = U_{k}(t)U_{k}(t-1)...U_{k}(1) \ket{\psi(0)},\nonumber\\
& =&\sum_{j=0}^{k-1}[\alpha_{0}(j,t)\ket{j_p,0_c} + \alpha_{1}(j,t)\ket{j_p,1_c}],
\end{eqnarray}
where, $\alpha_{0}(j,t)\text{ and } \alpha_{1}(j,t)$ are amplitudes for the states $\ket{j_p,0_c}$ and $\ket{j_p,1_c}$ respectively.
\noindent

\textit{Measuring Entanglement.---}
The initial quantum state in Eq.~(\ref{equ1}) is pure and separable, and it evolves unitarily via DTQW. We use entanglement entropy ($E$) to quantify the entanglement between the coin and position degrees of freedom of the time-evolved quantum state $\ket{\psi(t)}$~\cite{janzing_2009}. Let $\rho_{\psi}$ be density operator for $\ket{\psi(t)}$ i.e., $\rho_{\psi} = \ket{\psi(t)}\bra{\psi(t)}$ and reduced density operator ($\rho_c$) for the coin space is, $\rho_c \equiv \text{Tr}_p(\rho_{\psi})$, where the partial trace $\text{Tr}_p$ is taken over the position degrees of freedom. The eigenvalues of the reduced density matrix $\rho_c$ are,
$E_{\pm} = \frac{1}{2} \pm |\vec{n}|\;$,
\noindent
with,
$\vec{n} = \Big(\Re(\Sigma_j \alpha_{1}(j,t)\alpha_{2}^*(j,t)),\; \Im(\Sigma_j \alpha_{1}(j,t)\alpha_{2}^*(j,t)),\\
\frac{1}{2}\Sigma_j(|\alpha_{1}(j,t)|^2 - |\alpha_{2}(j,t)|^2)\Big).$
The entanglement entropy $E$ is the von-Neumann entropy for the coin state's reduced density matrix $\rho_c$. $E$ is defined as, $E(\rho_c)=$ $-$Tr($\rho_c$log$_{2}$$\rho_c$) with 0 for separable states and 1 for MESPS, and can be calculated via, $E = -E_{-}\text{log$_{2}$}E_{-}-E_{+}\text{log$_{2}$}E_{+} \;$.

To check whether our results are correct, we also calculate the Schmidt norm (another entanglement measure), which is given by,
$S = \sqrt{E_-} + \sqrt{E_+}\;,$
and for the present system with $\text{min}(\text{dim~}H_P,\text{dim~}H_C)=2$, $S$ for a MESPS is $\sqrt{2}$~\cite{gratsea_lewenstein_dauphin_2020,me-cb}. In Supplementary Material (SM) Sec.~\ref{App.B}, we show results from both the entanglement measures and their similar nature.

\textit{Periodicity of DTQW on cyclic graphs--}
Further, the QW on a $k$-cycle is said to be ordered or periodic if the walker reverts to its initial state after a time step, say $t=N$, irrespective of the initial quantum state. For an ordered QW with period $N$, we may write,
\begin{equation}
\ket{\psi(N)}=U_{k}(N)U_{k}(N-1)...U_{k}(1)\ket{\psi_i}=\ket{\psi_i}.
\label{equ6}
\end{equation}
If we apply the same coin at each time step in the above QW evolution, i.e., $U_{k}(t)=U_{k}(t-1)=...U_{k}(1)=U_{k}\text{(say)}$, then Eq.~(\ref{equ6}) is equivalent to,
$U^{N}_{k}\ket{\psi_i}=\sum^{2k}_{i=1}a_i\lambda^{N}_{i}\ket{\lambda_i},$ wherein the arbitrary $\ket{\psi_i}$ is expressed in terms of the eigenvalues $\{\lambda_i\}$ and eigenvectors $\{\ket{\lambda_i}\}$ of $U_{k}$, i.e., $\ket{\psi_i}=\sum^{2k}_{i=1}a_i\ket{\lambda_i}$.
From Eq.~(\ref{equ6}), the condition of periodicity for the QW follows:
$U_k^{N}=I_{2k} \;\text{or}\; \lambda_i^{N}=1, \;\;\forall\; i\in\{1,2,...,2k\}$. Any unitary evolution-operator which satisfies this condition gives a periodic probability distribution for the walker's position and yields ordered QW. Otherwise, the QW is said to be chaotic.
Furthermore, to simplify the problem of finding the eigenvalues of $U_k$ and hence the periodicity of the QW, the $2\cross2$ block circulant matrix $U_k$ is block diagonalized by using commensurate Fourier matrix tool as done in Ref.~\cite{cb-14}. Then the block diagonalized form of $U_k$ is given by $F_cU_{k}F_{c}^{\dagger}$$=\text{diag}[U_{k,0}, U_{k,1},..., U_{k,k-1}]$, wherein $F_c=F^{k}\otimes F^{2}$ with $F^{M}$ (with $M\in\{k,2\}$) being an $M\times M$ commensurate Fourier matrix, i.e., $F^{M}=(F^{M}_{m,n})=\frac{1}{\sqrt{M}}(e^{2\pi i\frac{mn}{M}})$ where $m,n=0,1,...,M-1$. The periodicity condition is satisfied if the eigenvalues $\lambda^{\pm}_{k,l}$ of each block $U_{k,l}$ satisfies the condition, $(\lambda^{\pm}_{k,l})^{\frac{N}{v}}=1$, where $v$ is the number of steps in the evolution which repeats itself in an evolution-operator sequence. In Refs.~\cite{cb-14,cb-13}, examples of parameter values for $U_k$ to obtain ordered QWs have been given. We discuss a unique analytical approach for obtaining values of such parameters, viz. $\{\rho,\gamma,\eta\}$ yielding recurrent MESPS via ordered QWs, with various evolution-operator sequences in \textit{Results}.


\begin{figure}[h]
\includegraphics[width = 8.8cm,height=4.5cm]{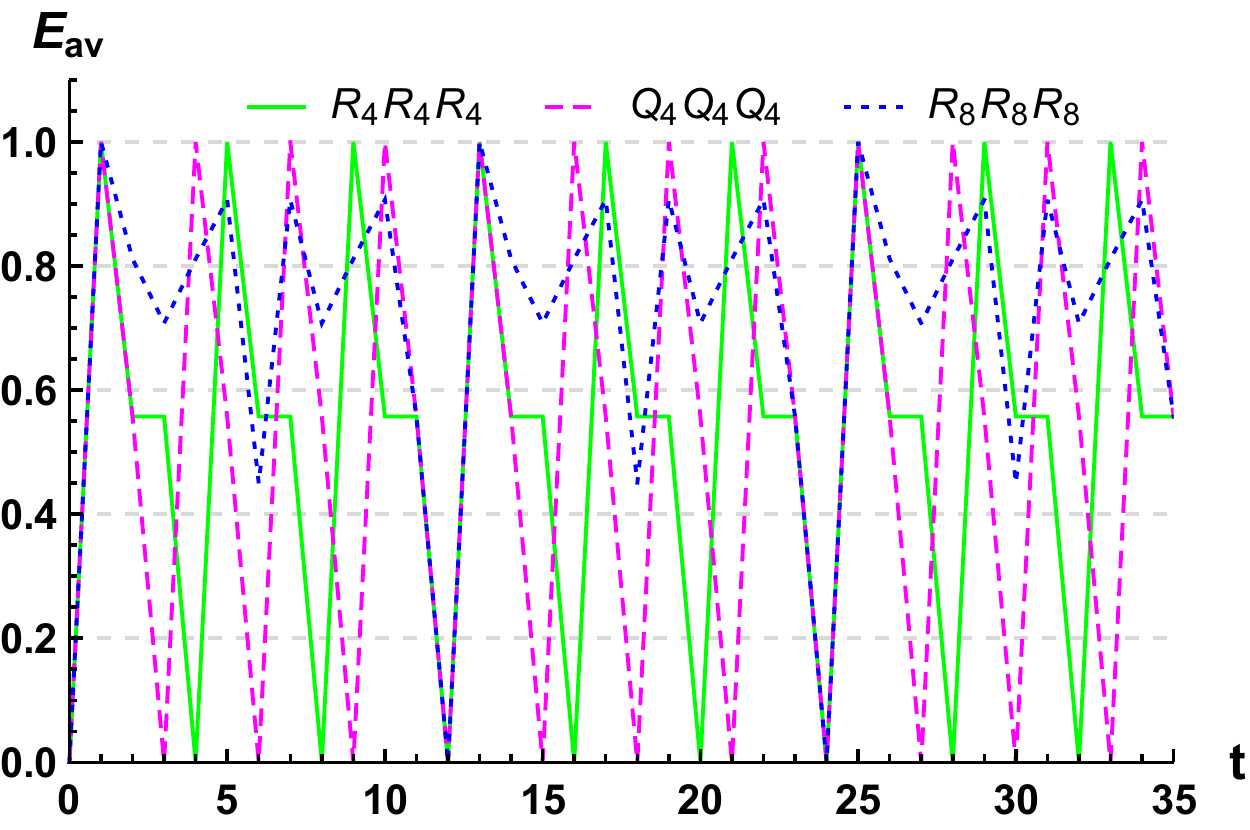}
\caption{$E_{av}$ versus time steps($t$) for single-coin evolution sequences: $R_4R_4R_4...$ (solid green), $Q_4Q_4Q_4...$ (dashed magenta) for 4-cycle and $R_8R_8R_8...$ (dotted blue) for 8-cycle, and an arbitrary separable initial state with $\phi=\frac{\pi}{6}$.}
\label{f2}
\end{figure}

\textit{Results:}
\textit{MESPS with single-coin evolution sequences.--}

A general framework for any single-coin $\hat{C}_{2}(\rho, \gamma, \eta)$ to yield MESPS at time step $t=1$ for the QW on any $k$-cycle via the single-coin evolution sequence $A_{k}A_{k}A_{k}...$ with evolution operator $A_{k}=U_{k}(\rho,\gamma,\eta)$ $= \hat{S}.[I_k\otimes \hat{C}_{2}(\rho,\gamma,\eta)]$, is established in SM Sec.~\ref{App.A}. A single coin of the form,
\begin{equation}
\hat{C}_{2}(\rho=\frac{1}{2}, \gamma, \eta)=
\frac{1}{\sqrt{2}}\begin{pmatrix}
1 & e^{i\gamma}\\
e^{i\eta} & -e^{i(\eta+\gamma)}
\end{pmatrix},
\label{equ7}
\end{equation}
under the constraint $(\gamma+\phi)\in\{\frac{\pi}{2},\frac{3\pi}{2}\}$,
generates MESPS at $t=1$ for any odd or even-cycle, or a line, from an arbitrary separable initial state Eq.~(\ref{equ1}). In addition, a subset of such arbitrary single coins, i.e., $\hat{C}_{2}(\rho=\frac{1}{2}, \gamma\in[0,\pi], \eta\in[0,\pi])$ with parameters $(\gamma+\eta)\in\{0,\frac{\pi}{2},\pi,\frac{3\pi}{2}\}$, yields recurrent or periodic MESPS (starting at time step $t=1$) on both 4-cycle and 8-cycle, see SM Sec.~\ref{App.A}.

Initial states (Eq.~(\ref{equ1})) having arbitrary $\phi\in[0,2\pi)$ values like $\phi$ = $\frac{\pi}{6},\frac{\pi}{5},\frac{\pi}{4},\frac{\pi}{3}, \frac{\pi}{2},\pi$, etc., can generate MESPS recurrently on both 4 and 8-cycles. For example, see Fig.~\ref{f2}, where the single-coin evolution sequence $R_{k}R_{k}R_{k}...$ with $R_{k}=U_{k}(\rho=\frac{1}{2},\gamma=\frac{\pi}{3},\eta=\frac{2\pi}{3})= \hat{S}.[I_k\otimes \hat{C}_{2}(\rho=\frac{1}{2},\gamma=\frac{\pi}{3},\eta=\frac{2\pi}{3})]$, yields recurrent MESPS on both $k=4$-cycle and $k=8$-cycle, for the initial state with $\phi=\frac{\pi}{6}$. Note that each data point in Fig.~\ref{f2} (and in the following figures) is an average of the entanglement entropy ($E_{av}$), and the average is taken over $\theta$ with the mentioned $\phi$ value and is evaluated as
$E_{av} = \frac{1}{\pi} \int_{0}^{\pi} E\;d\theta\; $. For MESPS, $E_{av}=1$.
The sequence $R_{4}R_{4}R_{4}...$ at $t=1,5,9,...$ yields MESPS on 4-cycle, with period 4, whereas the sequence $R_{8}R_{8}R_{8}...$ yields MESPS at $t=1,13,25,...$ (with period 12) on 8-cycle. Here, the coin $\hat{R}=\hat{C_2}(\frac{1}{2},\frac{\pi}{3},\frac{2\pi}{3})$ which is applied at each QW-time-step, is involutory i.e., $\hat{R}^2=I_2$. However, the use of involutory coins is not a necessary condition to generate recurrent MESPS; for instance, the single non-involutory coin evolution sequence $Q_4Q_4Q_4...$ with $Q_4=U_{4}(\frac{1}{2},\frac{\pi}{3},\frac{\pi}{6})$ (i.e., an non-involutory coin $\hat{Q}=\hat{C_2}(\rho=\frac{1}{2},\gamma=\frac{\pi}{3},\eta=\frac{\pi}{6})$ with $\hat{Q}^2\ne I_2$, applied at each time-step) on 4-cycle, yields MESPS with period 3 at $t=1,4,7,10...$ (see, Fig.~\ref{f2}), for the same initial state.

\begin{figure}[h]
\includegraphics[width = 8.8cm,height=4.5cm]{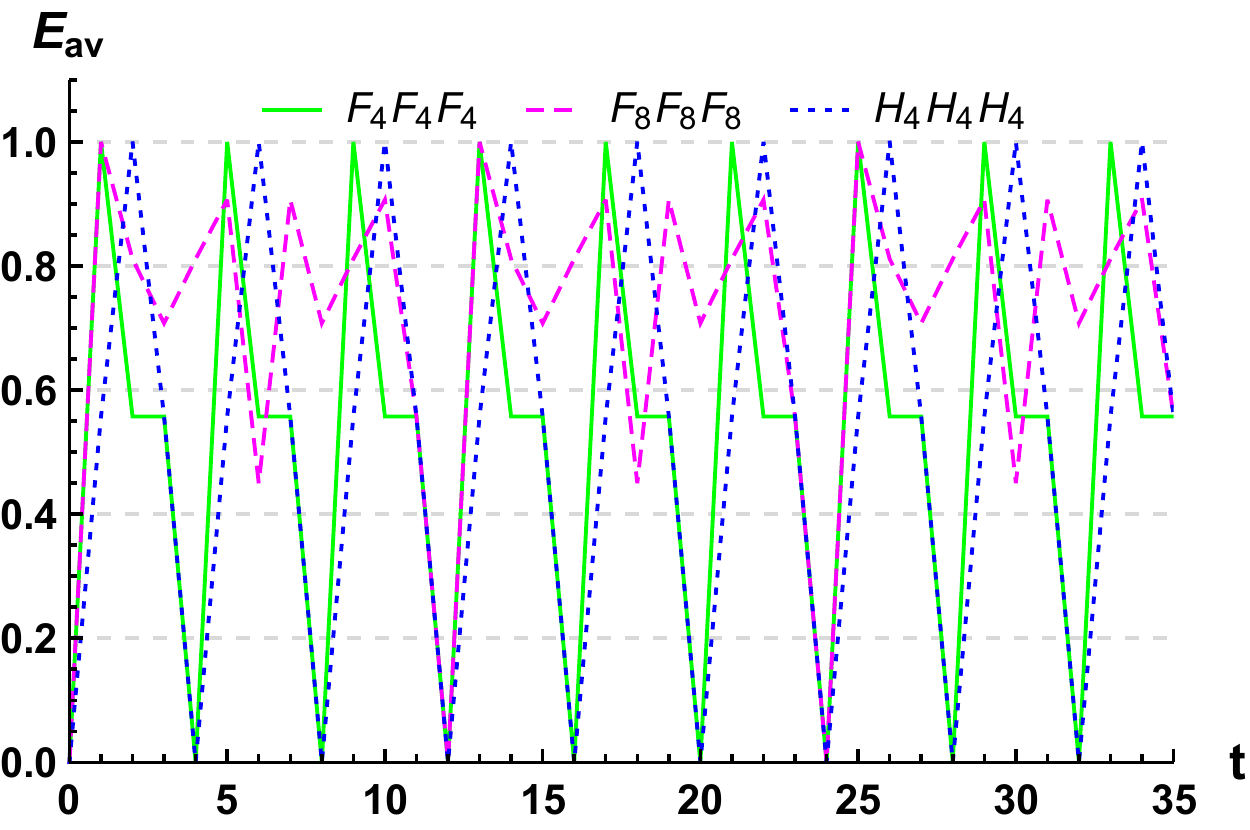}
\caption{ $E_{av}$ versus time steps($t$) with single non-involutory coin evolution sequences: $F_{4}F_{4}F_{4}...$ (solid green) for 4-cycle, $F_{8}F_{8}F_{8}...$ (dashed magenta) for 8-cycle, and single involutory coin evolution sequence $H_4H_4H_4...$ (dotted blue) for 4-cycle, for arbitrary separable initial state with $\phi=\pi$. }
\label{f3}
\end{figure}

By considering another separable initial state Eq.~(\ref{equ1}) with $\phi=\pi$, we find that non-involutory Fourier coin $\hat{F}=\hat{C}_{2}(\frac{1}{2}, \frac{\pi}{2}, \frac{\pi}{2})$, via its single-coin evolution sequence $F_{k}F_{k}F_{k}...$ with $F_k=U_k(\frac{1}{2},\frac{\pi}{2}, \frac{\pi}{2})$, yields recurrent MESPS on both $k=4$ and $k=8$-cycles, as shown in Fig.~\ref{f3}. The sequences $F_{4}F_{4}F_{4}...$ and $F_{8}F_{8}F_{8}...$ generate MESPS respectively at $t=1,5,9,13,...$ with period 4 and at $t=1,13,25,...$ with period 12. Again, for
$\phi=\pi$, the involutory single-coin evolution sequence $H_4H_4H_4...$ with $H_{4}=U_4(\frac{1}{2},0,0)$ (i.e., Hadamard coin $\hat{H}=\hat{C_2}(\rho=\frac{1}{2},\gamma=0,\eta=0)$ applied at each time-step) on 4-cycle, yields MESPS (with
period 4) at $t=2,6,10,...$ (since $\gamma+\phi=\pi$) as shown in Fig.~\ref{f3}.

\begin{figure}[h]
\includegraphics[width = 8.8cm,height=4.5cm]{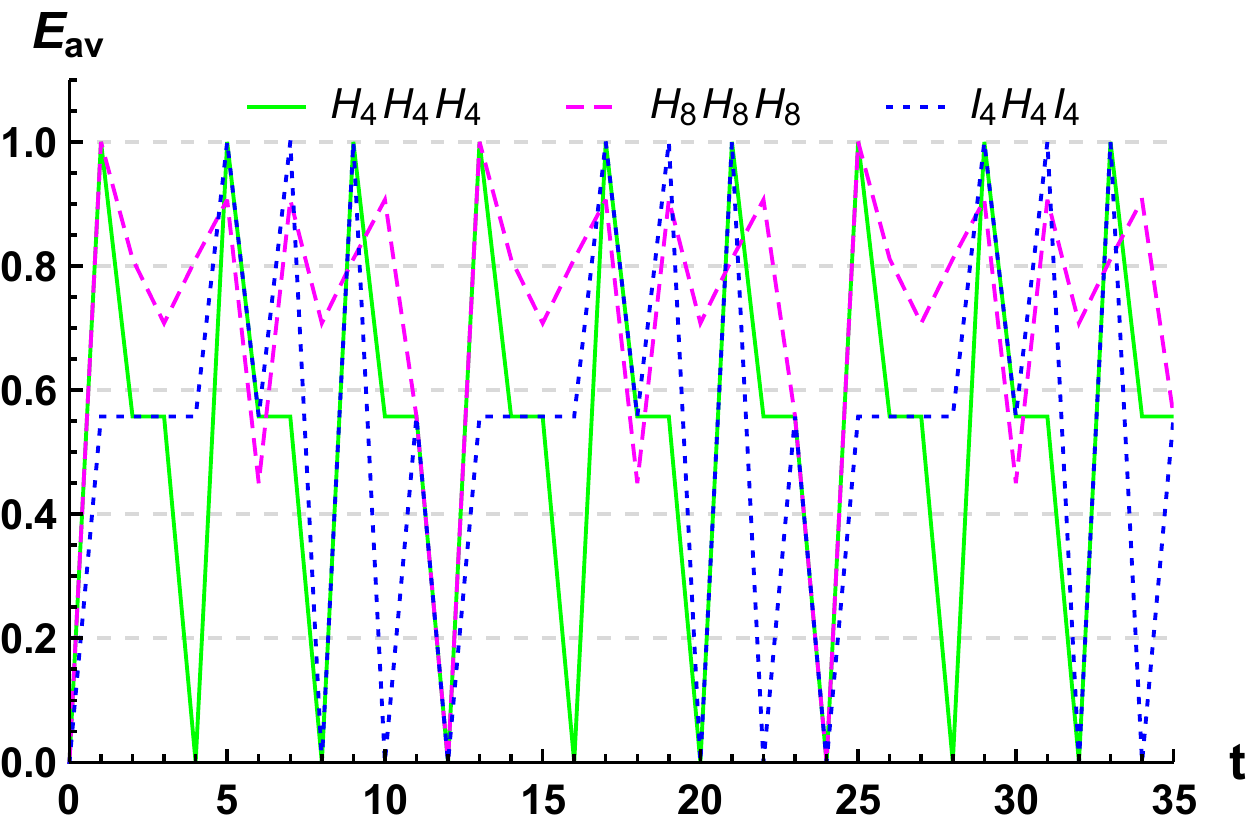}
\caption{$E_{av}$ versus time steps($t$) with evolution sequences $H_{4}H_{4}H_{4}...$ (solid green), $I_4H_4I_4...$ (dotted blue) for 4-cycle and $H_{8}H_{8}H_{8}...$ (dashed magenta) for 8-cycle, for arbitrary separable initial state with $\phi=\frac{\pi}{2}$.}
\label{f4}
\end{figure}

Furthermore, with separable initial state Eq.~(\ref{equ1}) having $\phi=\frac{\pi}{2}$, the sequence $H_{4}H_{4}H_{4}...$ on 4-cycle, yields MESPS at $t=1,5,9,...$ (here $ \gamma+\phi=\frac{\pi}{2}$) with period 4, as shown in Fig.~\ref{f4}. Similarly, sequence $H_{8}H_{8}H_{8}...$ with $H_{8}=U_8(\frac{1}{2},0,0)$ yields recurrent MESPS at $t=1,13,25,...$ with period 12 on 8-cycle for the same initial state, see Fig.~\ref{f4} (see more examples in SM Sec.~\ref{App.A} and~\ref{App.B}).
The periodic behavior of $H_{k}H_{k}H_{k}...$ in generating MESPS is supported by its ordered QW dynamics on both $k=4$ and $k=8$ cycles, see SM Sec.~\ref{App.B} for its analytical proof. Besides, we show that more than one MESPS can also occur within the period of the QW.

Unfortunately, for both $k=3$-cycle and $k=5$-cycle, we do not see periodic MESPS with single-coin evolution sequences. However, an arbitrary coin from Eq.~(\ref{equ7}) subject to the constraint $(\gamma+\phi)\in\{\frac{\pi}{2},\frac{3\pi}{2}\}$, yields MESPS at $t=1$ irrespective of whether it is an even or odd cycle, see SM Sec.~\ref{App.B}.

Note that a QW for a single-coin evolution sequence $A_{k}A_{k}A_{k}...$ is the simplest in terms of experimental setup as it just uses the same coin $\hat{C_2}$ (Eq.~(\ref{equ7})) at each time step~\cite{expt-cyclicQW}. In other words, the same setup will be sufficient for its realization. Thus, the above-established general framework using a single coin (Eq.~(\ref{equ7})) for recurrent generation of MESPS, is pivotal and resource-saving.


\textit{MESPS with effective-single or two-coin evolution sequences.--}
We execute several numerical experiments by forming multiple deterministic coin evolution sequences such as $A_{k}B_{k}A_{k}A_{k}B_{k}A_{k}...$, $A_{k}B_{k}A_{k}B_{k}...$, $ A_{k}B_{k}B_{k}A_{k}B_{k}B_{k}...$, $ A_{k}A_{k}B_{k}A_{k}A_{k}B_{k}...$, etc., where, $A_{k}=U_{k}(\rho,\gamma,\eta)$ $= \hat{S}.[I_k\otimes \hat{C}_{2}(\rho,\gamma,\eta)]\;$ and $B_{k}=U_{k}(\rho^{'},\gamma^{'},\eta^{'})$ $= \hat{S}.[I_k\otimes \hat{C}_{2}(\rho^{'},\gamma^{'},\eta^{'})]\;$. Here, we consider three coin operators (Eq.~(\ref{equ2})): Hadamard $\hat{H}$, Grover $\hat{X}=\hat{C}_{2}(\rho=0, \gamma=0, \eta =0)$, Identity $\hat{I}=\hat{C}_{2}(\rho=1,\gamma,\eta \ni \gamma+\eta=\pi)$. If $\hat{C}_2=\hat{X}$, we have evolution operator $X_k=U_{k}(0,0,0)$ $= \hat{S}.[I_k\otimes \hat{X}]\;$, and if $\hat{C}_2=\hat{I}$, then evolution operator $I_k=U_{k}(1,0,\pi)$ $= \hat{S}.[I_k\otimes \hat{I}]\;$. The primary idea behind such experiments was to reveal evolution-operator sequences involving either two coins such as $H_{k}H_{k}X_{k}...$, $H_{k}X_{k}H_{k}X_{k}...$, etc. or, effective-single coin (i.e., $I_{k}$ with either $H_{k}$ or $X_{k}$) such as $I_{k}H_{k}I_{k}..., H_{k}I_{k}I_{k}...$, etc., which yield recurrent MESPS. We first discuss effective single-coin evolution sequences and then the two-coin evolution sequences to generate MESPS via DTQWs on either even ($k=4$) or odd ($k\in\{3,5\}$)-cycle. Notably, the effective-single coin evolution sequences like $I_{k}H_{k}I_{k}...$ or $H_{k}I_{k}I_{k}...$ consists of a single coin (here $\hat{H}$) with Identity ($\hat{I}$) and their experimental implementation is resource-saving too as no extra device is required for Identity coin operation although it is slightly complex than single-coin implementation~\cite{fang}.

\begin{figure}[h]
\includegraphics[width=8.8cm,height=4.5cm]{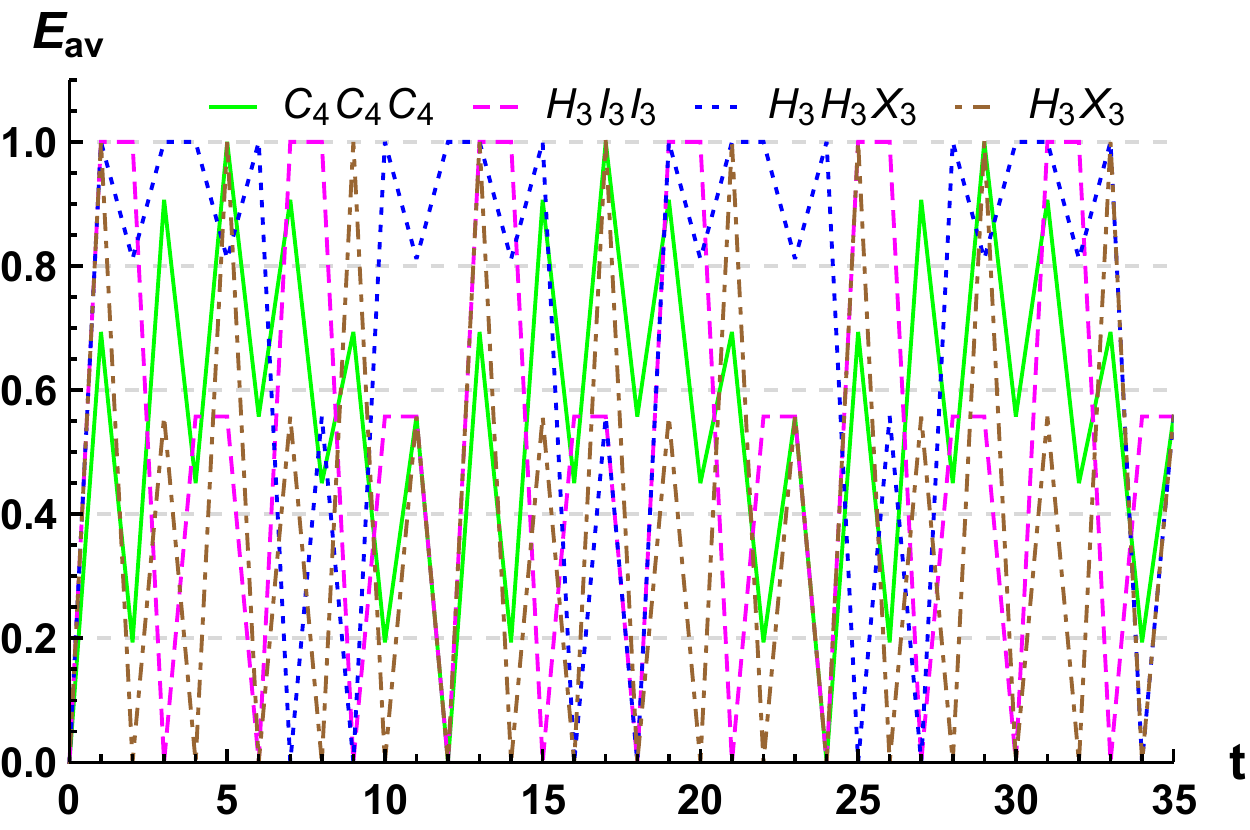}
\caption{$E_{av}$ versus time steps($t$) with evolution sequences: $C_{4}C_{4}C_{4}...$ (solid green) for 4-cycle, and $H_{3}I_{3}I_{3}...$ (dashed magenta), $H_{3}H_{3}X_{3}...$ (dotted blue), $H_{3}X_{3}...$ (dot-dashed brown), for 3-cycle, for an arbitrary separable initial state with $\phi=\frac{\pi}{2}$.}
\label{f5}
\end{figure}

We consider an arbitrary separable initial state Eq.~(\ref{equ1}) with $\phi=\frac{\pi}{2}$, and first discuss with the 4-cycle, the effective-single coin evolution sequences $I_{4}H_{4}I_{4}..., H_{4}I_{4}I_{4}... $ and $H_{4}I_{4}H_{4}I_{4}...$. We observe that the $E_{av}$ values generated via the sequence $I_{4}H_{4}I_{4}...$ follow a periodic trend, see Fig.~\ref{f4}. This observation is well supported by the periodic probability distribution $P(x=0)$ for the walker position at $\ket{0_p}$, in other words, the sequence $I_{4}H_{4}I_{4}...$ not only generates MESPS at $t=5,7,9,17...$ with period 12 but also an ordered QW, see SM Sec.~\ref{App.B}. Analytically one can also prove this by exploiting the periodicity condition, beginning with the eigenvalues of the $U_{4,1}$-block of the evolution operator $(U_{4})^{3}$, see Eq.~(\ref{equ3}),
$\lambda^{U_{4}U_{4}U_{4}}_{4,1}=\frac{1}{2}i\sqrt{\rho}e^{\frac{3}{2}i(\gamma+\eta)}(e^{-\frac{1}{2}i(\gamma+\eta)}+e^{\frac{1}{2}i(\gamma+\eta)})
(-3+2\rho+(e^{-i(\eta+\gamma)}+e^{i(\gamma+\eta)})\rho)\;.$ Similarly, the $U_{4,1}$-block's eigenvalues for the sequence $I_{4}H_{4}I_{4}$ give,
$\lambda^{I_{4}H_{4}I_{4}}_{4,1}= \frac{i}{\sqrt{2}}\;\;.$ Herein, $\lambda_{4,1}^{\tilde{U}}$ represents the sum, $\frac{\lambda_{4,1}^{+}+\lambda_{4,1}^{-}}{2}$, for the evolution  $\tilde{U}=(U_{4})^{3} $ or $I_{4}H_{4}I_{4}$.
Equating $\lambda^{U_{4}U_{4}U_{4}}_{4,1}$ with $\lambda^{I_{4}H_{4}I_{4}}_{4,1}$ for $(\gamma+\eta)=0$, we get $\rho= \frac{2+\sqrt{3}}{4}\;,$
which is an exact match with $\rho$ obtained in Ref.~\cite{cb-14} for a periodic QW with period $N=24$. With this description for $I_{4}H_{4}I_{4}...$ sequence giving an ordered QW, we observe that single involutory-coin evolution sequence $C_{4}C_{4}C_{4}...$ (i.e., coin $\hat{C}=\hat{C}_{2}(\rho=\frac{2+\sqrt{3}}{4},\gamma=0,\eta=0)$ applied at each time-step) generates MESPS with period 12 at $t=5,17,29,...$, see Fig.~\ref{f5}. It is another method besides Eq.~(\ref{equ7}) to obtain the condition for the single coin to give recurrent MESPS. Moreover, effective-single coin evolution sequences $H_{4}I_{4}I_{4}...$ and $H_{4}I_{4}H_{4}I_{4}...$ yield periodic MESPS with periods 12 and 4 at time steps $t=1,3,5,13,...$ and $t=1,5,9,...$ respectively (see SM Sec.~\ref{App.B}).

We also observe that two-coin evolution sequence $H_{4}H_{4}X_{4}...$ gives recurrent MESPS with period 6 at $t=1,3,7,9,13,...$ (proof of this periodicity is in SM Sec.~\ref{App.B}), whereas the sequence $H_{4}X_{4}H_{4}X_{4}...$ gives recurrent MESPS with period 4 at $t=1,5,9,...$, for 4-cycle.

Moving now to 3-cycle, the effective-single coin evolution sequence $H_{3}I_{3}I_{3}...$ yields periodic MESPS with period 6 at $t=1,2,7,8...$, but the sequence $I_{3}H_{3}I_{3}...$ renders ordered QWs without MESPS, whereas $H_{3}I_{3}H_{3}I_{3}...$ renders chaotic QW with MESPS at $t=1,2$, see Fig.~\ref{f5} and SM Sec.~\ref{App.B}. However, exploiting the periodicity condition for $H_{3}I_{3}I_{3}...$ sequence does not yield a MESPS-generating single-coin evolution sequence, unlike the case for $I_{4}H_{4}I_{4}...$ sequence.

From Fig.~\ref{f5}, we also observe that the two-coin evolution sequences $H_{3}H_{3}X_{3}...$ and $H_{3}X_{3}H_{3}X_{3}...$ generate recurrent MESPS respectively at $t=1,3,4,6,10,...$(with period 9) and $t=1,5,9,...$(with period 4) via the DTQW on the 3-cycle. For proof of this periodicity and results on 5-cycle yielding recurring MESPS via effective-single and two-coin evolution sequences, see SM Sec.~\ref{App.B}.


Interestingly, only employing $H_{k}H_{k}X_{k}...,\; H_{k}I_{k}I_{k}..., $ and $H_{k}X_{k}H_{k}X_{k}...$ on a ($k=3$)-cycle one can obtain MESPS at all time steps up to 10, whereas on a ($k=4$)-cycle these sequences give MESPS at all odd time steps $t\le10$, see Figs.~\ref{f5} and SM Sec.~\ref{App.B}. As these sequences also beget periodic QWs; thus, one obtains MESPS at larger time steps ($t>10$) as well. Moreover, on a ($k=5$)-cycle, just the sequences $H_{k}H_{k}X_{k}...$ and $H_{k}I_{k}I_{k}...$, generate MESPS at all time steps $t\le10$, see SM Sec.~\ref{App.B} and ~\ref{App.C}.

\textit{Cryptography protocol.--}
Periodic MESPS generation via our DTQW scheme can be exploited to design a quantum cryptographic protocol~\cite{cb-ap,crypt15}. Herein we put forth an example with the single-coin evolution sequence $H_{4}H_{4}H_{4}...$ for a 4-cycle (i.e., the Hadamard QW as shown in Fig.~\ref{f4}), to perform a secure encryption-decryption of a message with the following steps.

Step-1: Alice wants to send a message $m\in{\{0,1,2,3\}}$ to Bob. Bob forms the public key as $\ket{\psi_{pk}}=A\ket{j_p}\ket{q_c}$, where $A=(H_{4})^5$, $\ket{j_p}$ with $j\in\{0,1,2,3\}$ and $\ket{q_c}=\cos(\frac{\theta}{2})\ket{0_c}+i\sin(\frac{\theta}{2})\ket{1_c}$, with $\theta\in[0,\pi],\;\phi=\frac{\pi}{2}$, are respectively the position and coin states of the quantum walker. As shown in Fig.~\ref{f4}, $(H_{4})^{{4n+1}}$ with $n=0,1,2...$ can generate MESPS periodic in time, with $(H_{4})^8=I_8$. Thus, $\ket{\psi_{pk}}$ is a MESPS. After generating this MESPS $|\psi_{pk}\rangle$, which acts as the public key, Bob sends it to Alice.

Step-2 (Encryption): Alice encodes the message via: $\ket{\psi(m)}=(T_m\otimes I_c)\ket{\psi_{pk}},$ where $T_m=\sum^{3}_{i=0}\ket{((i+m)\text{ mod }4)_p}\bra{i_p},$ akin to the shift operator with $I_c=I_2$, and sends it to Bob.

Step-3 (Decryption): Bob then decrypts the message by operating $W=(H_{4})^{3}$ from which he gets $\ket{((j+m)\text{ mod }4)_p,q_c}$. Bob reads $m{'}=(j+m)\text{ mod}\;4$ from the position ket, and from which he securely obtains Alice's message $m$.

The security of this MESPS-based cryptographic protocol, i.e., resilience against any eavesdropper attack like man-in-the-middle, intersept-and-resend, etc.~\cite{qkd-attack2020}, is described in SM Sec.~\ref{App.D}.

\textit{Conclusions.---} This letter provides a novel scheme to generate MESPS from separable initial quantum-states via DTQWs on $k$-cycles with $k\in\{3,4,5,8\}$, with just a single coin and with both effective-single coin and two-coin evolution sequences. We established a general framework that predicts coins yielding MESPS at time step $t=1$ via QW on any $k$-cycle with single-coin evolution sequences from any arbitrary initial separable-state (with any $\phi$ value subject to certain constraints). A subset of the coins yields recurrent MESPS on both 4- and 8-cycles. An analytical proof for periodic QW which supports the recurrent generation of MESPS has been established, and more than one MESPS can occur within the period of the QW, see SM Sec.~\ref{App.A} and \ref{App.B}.

In addition, we show that with a 4-cycle, effective-single and two-coin evolution sequences (e.g., $I_{4}H_{4}I_{4}$, $H_{4}H_{4}X_{4}...,$, etc.) and single-coin evolution sequence $C_{4}C_{4}C_{4}...$ (obtained from $I_{4}H_{4}I_{4}$), individually yield recurrent MESPS, from
the initial separable-state with $\phi=\frac{\pi}{2}$. Finally, with effective-single and two-coin evolution sequences, we show recurrent MESPS generation (with the same initial state) on $3-$ and $5-$cycles. In the 3-cycle case, the sequences $H_{3}I_{3}I_{3}..., H_{3}H_{3}X_{3}...,$ and $H_{3}X_{3}H_{3}X_{3}...$ altogether give MESPS at all $t\le10$, whereas in the 5-cycle case, with sequences $H_{5}H_{5}X_{5}...$ and $ H_{5}I_{5}I_{5}...$, one can obtain MESPS at all $t\le10$. In SM Sec.~\ref{App.C}, we summarize the evolution sequences to generate MESPS at time steps up to 10 and beyond with the cyclic graphs.

We have also outlined the steps to implement our scheme in quantum cryptography. One can experimentally implement our proposed scheme using linear optical elements such as half-wave plates (HWPs), quarter-wave plates (QWPs), and polarizing beam splitters (PBSs), along with a fast switching electro-optical modulator (EOM), wherein the photon's polarization degree of freedom encodes the coin state with the position state is encoded into different time bins of the photon~\cite{expt-cyclicQW,vieira_amorim_rigolin_2013}. Evaluating the entanglement entropy requires post-processing measurements like average polarizations of the photon by proper arrangement of an HWP and QWP~\cite{vieira_amorim_rigolin_2013,avgpol}.

A comparison of our work with other relevant works (DTQWs on 1D line)~\cite{r_zhang2022,gratsea2020universal,me-cb, fang,gratsea_lewenstein_dauphin_2020} can be found in SM Sec.~\ref{App.E}. Apart from opening a unique avenue for MESPS generation, our letter significantly outperforms other schemes in model simplicity and resource-saving architecture and periodically yields MESPS at both small and large time-steps. We provide a Python code for numerical experiments in SM Sec.~\ref{App.F}.

Our presented work will significantly contribute towards state-of-art controlled (maximal) entanglement generation protocols, a fundamental resource in quantum computing, teleportation, and cryptography and a prerequisite for quantum-information-processing tasks.

\textit{Acknowledgement.---} Colin Benjamin would like to thank Science and Engineering Research Board (SERB) for funding under the Core Research grant "Josephson junctions with strained Dirac materials and their application in quantum information processing," Grant No. CRG/2019/006258.


\twocolumngrid

\newpage
\onecolumngrid

\hspace{5cm}\underline{\textbf{\large SUPPLEMENTARY MATERIAL}}
\vspace{0.5cm}

Here in Sec.~\ref{App.A}, we provide a framework for arbitrary unitary coins that yield MESPS (maximally entangled single-particle states) at the first time step of the QW (quantum walk) via single coin evolution on any
$k$-cycle, and also for coins that yield recurrent MESPS on both 4- and 8-cycles via their single-coin evolution sequence from an arbitrary separable initial state. In Sec.~\ref{App.B}, we provide more details of our results, including the generation of MESPS and non-maximal single particle entanglement (SPE) via single-coin, effective-single coin, and two-coin evolution sequences. The proof for the periodicity in QW dynamics and reasons for MESPS periodicity via single-coin evolution sequences is also provided. We also show the occurrence of recurrent MESPS with periodic QW via effective single and two-coin evolution sequences. We juxtaposed evolution sequences yielding MESPS generated from an arbitrary separable initial state with $k$-cycles ($k\in\{3,4,5,8\}$) in Sec.~\ref{App.C}. The security of our MESPS-based quantum cryptographic scheme is discussed in Sec.~\ref{App.D}. We compare our work with other relevant works in Sec.~\ref{App.E}. Finally, we provide a Python code to generate figures of the letter in Sec.~\ref{App.F} for interested researchers.

\section{Condition for generating MESPS via Single coin}
\label{App.A}
As observed in the main text (Figs.~2-4), some single-coin evolution sequences such as $H_4H_4H_4...,\; R_4R_4R_4...$ on 4-cycle, yield MESPS at time step $t=1$ of the QW, from the separable initial state,
\begin{equation}
\ket{\psi(t=0)} = \cos(\frac{\theta}{2})\ket{0_p,0_c} + e^{i\phi}\sin(\frac{\theta}{2})\ket{0_p,1_c}.
\label{eq1}
\end{equation}
The general initial state in Eq.~(\ref{eq1}) is the same as the state mentioned in Eq.~(1) of the main manuscript. This exciting result leads us to find a condition for any single coin that could lead to MESPS at time step $t=1$ on any $k$-cycle (i.e., a cyclic graph with $k$ sites with $k\in\{3,4,5...\}$), from an arbitrary separable initial state, i.e., with any $\phi$ value (Eq.~(\ref{eq1})). To do so, let us consider the arbitrary coin operator as in the main text, i.e.,
\begin{equation}
\hat{C}_{2}(\rho, \gamma, \eta) =
\begin{pmatrix}
\sqrt{\rho} & \sqrt{1-\rho}e^{i\gamma}\\
\sqrt{1-\rho}e^{i\eta} & -\sqrt{\rho}e^{i(\gamma+\eta)}
\end{pmatrix},
\label{eq2}
\end{equation}
where $0\leq\rho\leq1$ \text{and} $0\leq\gamma,\eta\leq\pi$, and using the Schmidt norm ($S_{av}$) or von-Neumann entropy ($E_{av}$) to measure the generated entanglement which is 1 for MESPS.

At time step $t=1$ via the single-coin QW, we get the quantum state,
\begin{equation}
\ket{\psi(1)} = \nu\{\cos(\frac{\theta}{2})+\mu e^{i(\gamma+\phi)} \sin(\frac{\theta}{2})\}\ket{(k-1)_p,0_c} +\nu\{\mu e^{i\eta}\cos(\frac{\theta}{2})- e^{i(\gamma+\phi+\eta)} \sin(\frac{\theta}{2})\}\ket{1_p,1_c},
\label{eq3}
\end{equation}
where $\nu=\sqrt{\rho}$ and $\mu=\sqrt{\frac{1-\rho}{\rho}}$.

Then, the Schmidt norm $S$ (for the procedure of calculating Schmidt norm, see main text page~2),
\begin{equation}
S=\nu \cos(\frac{\theta}{2})\big\{\sqrt{\mu^2+\tan^2(\frac{\theta}{2})-2\mu\cos(\gamma+\phi)\tan(\frac{\theta}{2})}+\sqrt{1+\mu^2\tan^2(\frac{\theta}{2})+2\mu\cos(\gamma+\phi)\tan(\frac{\theta}{2})}\;\big\}
\label{eq4}
\end{equation}

Now, for $(\gamma+\phi)=\frac{\pi}{2}$ or $\frac{3\pi}{2}$, we get average Schmidt norm,
\begin{equation}
S_{av}= \frac{1}{\pi} \int_{0}^{\pi}d\theta \; \; \frac{S}{\sqrt{2}} = \frac{\nu}{\sqrt{2}\pi}\int_{0}^{\pi}d\theta \; \; \cos(\frac{\theta}{2})\{\sqrt{\mu^2+\tan^2(\frac{\theta}{2})}+\sqrt{1+\mu^2 \tan^2(\frac{\theta}{2})}\;\}=\frac{2\sqrt{2}\nu}{\pi}E_I(1-\mu^2),
\label{eq5}
\end{equation}
where, $E_I(1-\mu^2)$ is a complete elliptic integral. We use the Schmidt norm to quantify the entanglement of MESPS as it is slightly easier to integrate.

Here, $S_{av}=1\implies\rho=\frac{1}{2}$ for arbitrary $\eta$ values. It implies that a general coin of the form,
\begin{equation}
\hat{C}_{2}(\rho=\frac{1}{2}, \gamma\in[0,\pi], \eta\in[0,\pi]) =
\frac{1}{\sqrt{2}}\begin{pmatrix}
1 & e^{i\gamma}\\
e^{i\eta} & -e^{i(\eta+\gamma)}
\end{pmatrix},
\label{eq6}
\end{equation}
with $(\gamma+\phi) \in \{\frac{\pi}{2},\frac{3\pi}{2}\}$,
yields MESPS at $t=1$ on any odd or even $k$-cycle.

\begin{figure}[H]
\centering
\includegraphics[width=1\linewidth]{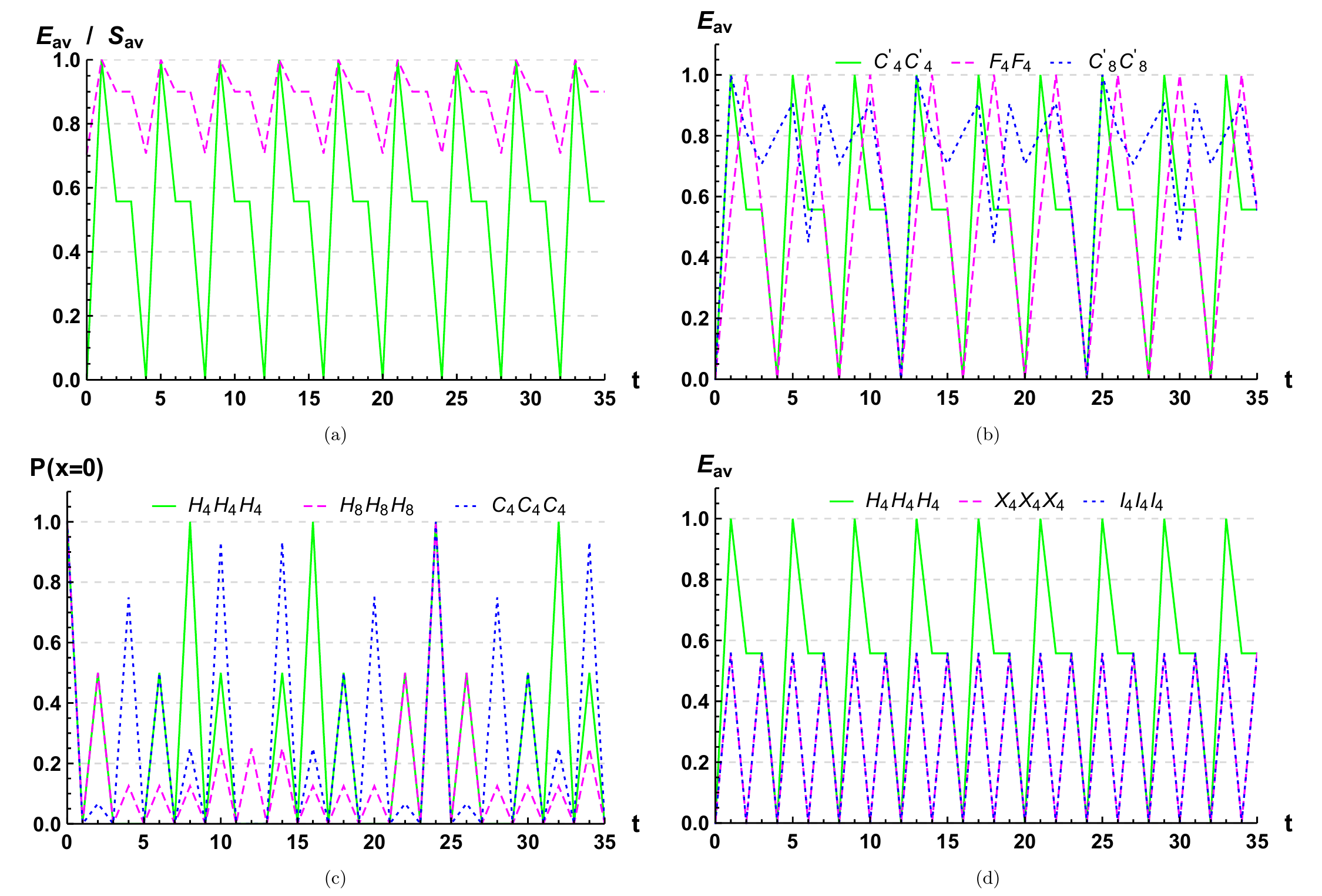}
\caption{(a) Average entanglement entropy $E_{av}$ (solid green) and average Schmidt norm $S_{av}$ (dashed magenta) versus time steps ($t$) with single-coin evolution sequence $H_{4}H_{4}H_{4}...$, for 4-cycle for the general separable initial state (Eq.~(\ref{eq1})) with $\phi=\frac{\pi}{2}$. (b) $E_{av}$ versus time steps($t$) with single non-involutory coin evolution sequences $C^{'}_{4}C^{'}_{4}C^{'}_{4}...$ (solid green), $F_{4}F_{4}F_{4}...$ (dashed magenta) for 4-cycle and $C^{'}_{8}C^{'}_{8}C^{'}_{8}...$ (dotted blue) for 8-cycle, for arbitrary initial state Eq.~(\ref{eq1}) with $\phi=\frac{\pi}{2}$. (c) Probability $P(x=0)$ of finding the walker at position $\ket{0_p}$ as function of time steps($t$) with the single coin evolution sequences $H_{4}H_{4}H_{4}...$ (solid green), $C_{4}C_{4}C_{4}...$ (dotted blue) for 4-cycle, and $H_{8}H_{8}H_{8}...$ (dashed magenta) for 8-cycle, with $\theta=0$ and $\phi=\frac{\pi}{2}$. (d) $E_{av}$ as function of time steps($t$) with single coin evolution sequences: $H_{4}H_{4}H_{4}...$ (solid green), $ X_{4}X_{4}X_{4}...$ 
 (dashed magenta), $I_{4}I_{4}I_{4}...$ (dotted blue), for 4-cycle with $\phi=\frac{\pi}{2}$.}
\label{sf1}
\end{figure}

For instance, from the separable initial state with $\phi=\frac{\pi}{2}$, the Hadamard coin $\hat{H}$ which is $\hat{C}_{2}(\rho=\frac{1}{2}, \gamma=0, \eta=0)$, yields MESPS at $t=1$ via the single-coin evolution QW on any $k$-cycle. It also generates MESPS at $t=1,5,9,13...$ with period 4 on a 4-cycle, see Fig.~\ref{sf1}(a). Similarly, the non-involutory coin $\hat{C^{'}}=\hat{C}_{2}(\rho=\frac{1}{2}, \gamma=\pi, \eta=0)=
\frac{1}{\sqrt{2}}\begin{pmatrix}
1 & -1\\
1 & 1
\end{pmatrix}$ with $\hat{C^{'}}^{2}\ne I_2$, yields MESPS at $t=1$ on any $k$-cycle. Particularly, it yields MESPS at $t=1,5,9...$ with period 4 on a 4-cycle, see Fig.~\ref{sf1}(b).

This formalism enables one to predict the kind of coins that generate MESPS at $t=1$ via their single-coin evolution on a $k$-cycle. Additionally, using the QW-periodicity condition involving the Commensurate Fourier matrix, as mentioned in the main text, one can predict which of the general coins gives rise to recurrent or periodic MESPS. For example, with a 4-cycle (or 8-cycle) and $\phi=\frac{\pi}{2}$, the general coin $\hat{C}_{2}(\rho=\frac{1}{2}, \gamma\in\{0,\pi\}, \eta\in[0,\pi]) =
\frac{1}{\sqrt{2}}\begin{pmatrix}
1 & e^{i\gamma}\\
e^{i\eta} & -e^{i(\eta+\gamma)}
\end{pmatrix},$ where $(\gamma+\eta)=0,\pi,\frac{\pi}{2},\frac{3\pi}{2}$, yields ordered QWs (as shown by Ref.~\cite{cb-14}) and hence recurrent MESPS with different periods. For instance, the single coin evolution sequence $H_4H_4H_4...$ with Hadamard coin $\hat{H}=\hat{C}_{2}(\rho=\frac{1}{2}, \gamma=0, \eta=0)$, yields recurrent MESPS at $t=1,5,9,...$ with period 4, see Fig.~\ref{sf1}(a). Similarly, the non-involutory coin $\hat{C^{'}}=\hat{C}_{2}(\rho=\frac{1}{2}, \gamma=\pi, \eta=0)$, yields recurrent MESPS at $t=1,5,9,13...$ on a 4-cycle (see Fig.~\ref{sf1}(b)). Moreover, with $\phi=\frac{\pi}{6}$, the single coin evolution sequence $R_kR_kR_k...$ with the coin operator $\hat{R}=\hat{C}_{2}(\rho=\frac{1}{2}, \gamma=\frac{\pi}{3}, \eta=\frac{2\pi}{3})$ yields recurrent (periodic) MESPS on both $k=4$ and $k=8$-cycles, as shown in main letter Fig.~2.

Furthermore, a similar general discussion can be made for the initial separable state (Eq.~(\ref{eq1})) having $\phi=\pi$. This implies that a general coin of the form $\hat{C}_{2}(\rho=\frac{1}{2}, \gamma=\pm\frac{\pi}{2}, \eta\in[0,\pi]) =
\frac{1}{\sqrt{2}}\begin{pmatrix}
1 & \pm i\\
e^{i\eta} & \mp ie^{i\eta}
\end{pmatrix},$ yields MESPS at $t=1$ via the single-coin evolution QW on any $k$-cycle. Fourier coin (i.e., $\hat{C}_{2}(\rho=\frac{1}{2}, \gamma=\frac{\pi}{2}, \eta=\frac{\pi}{2})$) yielding MESPS at $t=1,5,9,...$ with period 4 on a 4-cycle, is an example of this case; see main letter Fig.~3.

Though the single coin, as in Eq.~(\ref{eq6}), yields MESPS at $t=1$ on $k\in\{3,5\}$-cycles via its single-coin evolution QW, we did not find any coin yielding periodic MESPS via the single-coin evolution sequence, using steps Eqs.~(\ref{eq1})-(\ref{eq5}) on these odd cycles.


\section{Recurrent MESPS and periodicity of quantum walks}
\label{App.B}
\subsection{\large 4-cycle and 8-cycle}
\subsubsection{Single-coin evolution sequences}
Fig.~\ref{sf1}(a) above shows the entanglement entropy and Schmidt norm values as functions of time steps($t$), generated by the single coin evolution sequence $H_{4}H_{4}H_{4}...$ on the 4-cycle, wherein each data point is an average of the Schmidt norm ($S_{av}$) or the entanglement entropy ($E_{av}$), and the average is taken over $\theta$ with fixed $\phi=\frac{\pi}{2}$ and is evaluated as,
$S_{av}=\langle\frac{S}{\sqrt{2}}\rangle = \frac{1}{\pi} \int_{0}^{\pi}d\theta \; \; \frac{S}{\sqrt{2}} \;,$
or
$
E_{av}=\langle E \rangle = \frac{1}{\pi} \int_{0}^{\pi} E\;d\theta \;.$
Since both give identical results for MESPS, i.e., $S_{av}=E_{av}=1$, we only calculate $E_{av}$ in this Letter. One can also observe that the sequence $H_{4}H_{4}H_{4}...$ generates MESPS at time steps $t=1,5,9,...$ with period 4 (via any of the entanglement measures). The ordered QW dynamics support this periodicity, as shown in Fig.~\ref{sf1}(c), which shows periodic probability distribution $P(x=0)$ for the walker position at $\ket{0_p}$ with sequence $H_{4}H_{4}H_{4}...$ for 4-cycle, along with those via sequences $H_{8}H_{8}H_{8}...$ for 8-cycle and $C_{4}C_{4}C_{4}...$ for 4-cycle. The single coin sequences $X_4X_4X_4...$ and $I_4I_4I_4...$ do not yield MESPS like the $H_4H_4H_4$, see Fig.~\ref{sf1}(d). The sequence $H_{4}H_{4}H_{4}...$ (i.e., Hadamard coin $\hat{C}_{2}(\rho=\frac{1}{2}, \gamma=0, \eta=0)$) applied at each QW time step) yield periodic QW with period 8, on 4-cycle, see Ref.~\cite{cb-14}.

The periodicity in DTQW indeed supports the periodicity of MESPS, but the periodicity of MESPS is not just due to the periodicity of DTQW. It can be proved analytically by going over the quantum states $\ket{\psi(t)}$ generated at time steps $t = 0 $ to $t = 9$ via sequence $H_{4}H_{4}H_{4}...$ in 4-cycle as follows:
\begin{align}
\label{eqn:eq7}
\begin{split}
\ket{\psi(t=0)} &=\ket{0_p}\otimes[(\alpha+\beta)\ket{0_c}+(\alpha -\beta)\ket{1_c}] \longrightarrow \text{Separable } (E_{av}=0),
\\
\ket{\psi(1)} &=\sqrt{2}\alpha \ket{3_p,0_c}+\sqrt{2}\beta \ket{1_p,1_c} \longrightarrow \text{MESPS }(E_{av}=1),
\\
\ket{\psi(2)} &=\alpha \ket{0_p,1_c}+\beta \ket{0_p,0_c}+\alpha \ket{2_p,0_c}-\beta \ket{2_p,1_c} \longrightarrow\text{Partially entangled } (E_{av}=0.557),
\\
\ket{\psi(3)} &=\frac{1}{\sqrt{2}}[(\alpha -\beta)(\ket{1_p,0_c}-\ket{1_p,1_c})+(\alpha +\beta) (\ket{3_p,0_c}+ \ket{3_p,1_c})] \longrightarrow\text{Partially entangled} (E_{av}=0.557),
\\
\ket{\psi(4)} &=(\alpha -\beta)\ket{2_p,1_c}+(\alpha +\beta) \ket{2_p,0_c} \longrightarrow\text{Separable } (E_{av}=0),
\\
\ket{\psi(5)} &=\sqrt{2}\alpha \ket{1_p,0_c}+\sqrt{2}\beta \ket{3_p,1_c} \longrightarrow \text{MESPS }(E_{av}=1),
\\
\ket{\psi(6)} &=\alpha \ket{0_p,0_c}-\beta \ket{0_p,1_c}+\alpha \ket{2_p,1_c}+\beta \ket{2_p,0_c} \longrightarrow \text{Partially entangled } (E_{av}=0.557),
\\
\ket{\psi(7)} &=\frac{1}{\sqrt{2}}[(\alpha -\beta)(\ket{3_p,0_c}-\ket{3_p,1_c})+(\alpha +\beta) (\ket{1_p,0_c}+\ket{1_p,1_c})] \longrightarrow \text{Partially entangled} (E_{av}=0.557),
\\
\ket{\psi(8)} &=(\alpha -\beta)\ket{0_p,1_c}+(\alpha+\beta)\ket{0_p,0_c} \longrightarrow\text{Separable } (E_{av}=0),
\\
\ket{\psi(9)} &=\sqrt{2}\alpha \ket{3_p,0_c}+\sqrt{2}\beta \ket{1_p,1_c} =\ket{\psi(1)} \longrightarrow \text{MESPS }(E_{av}=1),
\end{split}
\end{align}
where, $\alpha=\frac{1}{2}(\cos(\frac{\theta}{2})+e^{i\phi}\sin(\frac{\theta}{2}))$ and $\beta=\frac{1}{2}(\cos(\frac{\theta}{2})-e^{i\phi}\sin(\frac{\theta}{2}))$ with $\phi=\pi/2$.

Clearly, MESPS $\ket{\psi(1)}=\ket{\psi(9)}$ is supported by the ordered DTQW of period $8$ generated by the single-coin sequence $H_{4}H_{4}H_{4}...$ on 4-cycle, see Figs.~\ref{sf1}(a) and (c). But $\ket{\psi(5)}$ also leads to a MESPS like $\ket{\psi(1)}$ or $\ket{\psi(9)}$. It proves that more than one MESPS can occur within the period of the DTQW. However, the MESPS states may not be the same, but all have maximal entanglement.

Similarly, for the single coin evolution sequence $C^{'}_{4}C^{'}_{4}C^{'}_{4}...$ with the non-involutory coin $\hat{C^{'}}=\hat{C_2}(\rho=1/2,\gamma=\pi,\eta=0)$, we get MESPS with period 4 at $t=1,5,9,...$, see Fig.~\ref{sf1}(b). The quantum states for $C^{'}_{4}C^{'}_{4}C^{'}_{4}...$ sequence, up to $t=9$ are:
\begin{align}
\label{eqn:eq8}
\begin{split}
\ket{\psi(t=0)} &=\ket{0_p}\otimes[(\alpha+\beta)\ket{0_c}+(\alpha -\beta)\ket{1_c}] \longrightarrow \text{Separable } (E_{av}=0),
\\
\ket{\psi(1)} &=\sqrt{2}\alpha \ket{1_p,1_c}+\sqrt{2}\beta \ket{3_p,0_c} \longrightarrow \text{MESPS }(E_{av}=1),
\\
\ket{\psi(2)} &=\alpha \ket{2_p,1_c}+\beta \ket{2_p,0_c}-\alpha \ket{0_p,0_c}+\beta \ket{0_p,1_c} \longrightarrow\text{Partially entangled } (E_{av}=0.557),
\\
\ket{\psi(3)} &=\frac{1}{\sqrt{2}}[(\beta-\alpha)(\ket{1_p,0_c}+\ket{1_p,1_c})+(\alpha +\beta) (\ket{3_p,1_c}-\ket{3_p,0_c})] \longrightarrow\text{Partially entangled} (E_{av}=0.557),
\\
\ket{\psi(4)} &=(\beta-\alpha)\ket{2_p,1_c}-(\alpha +\beta) \ket{2_p,0_c} \longrightarrow\text{Separable } (E_{av}=0),
\\
\ket{\psi(5)} &=-\sqrt{2}\alpha \ket{3_p,1_c}-\sqrt{2}\beta \ket{1_p,0_c} \longrightarrow \text{MESPS }(E_{av}=1),
\\
\ket{\psi(6)} &=\alpha \ket{2_p,0_c}-\beta \ket{2_p,1_c}-\alpha \ket{0_p,1_c}-\beta \ket{0_p,0_c} \longrightarrow \text{Partially entangled } (E_{av}=0.557),
\\
\ket{\psi(7)} &=\frac{1}{\sqrt{2}}[(\alpha -\beta)(\ket{3_p,0_c}+\ket{3_p,1_c})+(\alpha +\beta) (\ket{1_p,0_c}-\ket{1_p,1_c})] \longrightarrow \text{Partially entangled} (E_{av}=0.557),
\\
\ket{\psi(8)} &=(\alpha -\beta)\ket{0_p,1_c}+(\alpha+\beta)\ket{0_p,0_c} \longrightarrow\text{Separable } (E_{av}=0),
\\
\ket{\psi(9)} &=\sqrt{2}\alpha \ket{1_p,1_c}+\sqrt{2}\beta \ket{3_p,0_c} =\ket{\psi(1)} \longrightarrow \text{MESPS }(E_{av}=1).
\end{split}
\end{align}

Herein too, $\alpha=\frac{1}{2}(\cos(\frac{\theta}{2})+e^{i\phi}\sin(\frac{\theta}{2}))$ and $\beta=\frac{1}{2}(\cos(\frac{\theta}{2})-e^{i\phi} \sin(\frac{\theta}{2}))$ with $\phi=\pi/2$.
Again, MESPS $\ket{\psi(1)}=\ket{\psi(9)}$
and $\ket{\psi(5)}$ is a MESPS too. Thus, QW via the sequence $C^{'}_{4}C^{'}_{4}C^{'}_{4}...$ is periodic with period 8, whereas MESPS generated via the sequence is periodic with period 4.

Fig.~\ref{sf1}(b) above shows periodic MESPS generated via the single non-involutory coin evolution sequences $C^{'}_{4}C^{'}_{4}C^{'}_{4}...$, $F_{4}F_{4}F_{4}...$ on 4-cycle; $C^{'}_{8}C^{'}_{8}C^{'}_{8}...$ on 8-cycle, for the general initial state (Eq.~(1) of the Letter) with $\phi=\frac{\pi}{2}$.

In the main text Fig.~3, we observe that the single-coin evolution sequence $H_{4}H_{4}H_{4}...$ on 4-cycle yields MESPS at $t=2,6,10,...$ with period 4, for the separable initial state with $\phi=\pi$. For this sequence, the quantum states for up to $t=9$ are as follows:
\begin{align}
\label{eqn:eq9}
\begin{split}
\ket{\psi(t=0)} &=\ket{0_p}\otimes[(\alpha^{'}+\beta^{'})\ket{0_c}+(\alpha^{'} -\beta^{'})\ket{1_c}] \longrightarrow \text{Separable } (E_{av}=0),
\\
\ket{\psi(1)} &=\sqrt{2}\beta^{'} \ket{1_p,1_c}+\sqrt{2}\alpha^{'} \ket{3_p,0_c} \longrightarrow \text{Partially entangled }(E_{av}=0.557),
\\
\ket{\psi(2)} &=\alpha^{'} \ket{2_p,0_c}-\beta^{'}\ket{2_p,1_c} +\alpha^{'}\ket{0_p,1_c} +\beta^{'} \ket{0_p,0_c} \longrightarrow\text{MESPS } (E_{av}=1),
\\
\ket{\psi(3)} &=\frac{1}{\sqrt{2}}[(\alpha^{'}-\beta^{'})(\ket{1_p,0_c}-\ket{1_p,1_c})+(\alpha^{'} +\beta^{'}) (\ket{3_p,1_c}+\ket{3_p,0_c})] \longrightarrow\text{Partially entangled} (E_{av}=0.557),
\\
\ket{\psi(4)} &=(\alpha^{'}-\beta^{'})\ket{2_p,1_c}+(\alpha^{'} +\beta^{'}) \ket{2_p,0_c} \longrightarrow\text{Separable } (E_{av}=0),
\\
\ket{\psi(5)} &=\sqrt{2}\alpha^{'}\ket{1_p,0_c} +\sqrt{2}\beta^{'}\ket{3_p,1_c} \longrightarrow \text{Partially entangled } (E_{av}=0.557),
\\
\ket{\psi(6)} &=\beta^{'} \ket{2_p,0_c}+ \alpha^{'}\ket{2_p,1_c}-\beta^{'} \ket{0_p,1_c}+\alpha^{'}\ket{0_p,0_c} \longrightarrow \text{MESPS }(E_{av}=1) ,
\\
\ket{\psi(7)} &=\frac{1}{\sqrt{2}}[(\alpha^{'} -\beta^{'})(\ket{3_p,0_c}-\ket{3_p,1_c})+(\alpha^{'} +\beta^{'}) (\ket{1_p,0_c}+\ket{1_p,1_c})] \longrightarrow \text{Partially entangled} (E_{av}=0.557),
\\
\ket{\psi(8)} &=(\alpha^{'} -\beta^{'})\ket{0_p,1_c}+(\alpha^{'}+\beta^{'})\ket{0_p,0_c} \longrightarrow\text{Separable } (E_{av}=0),
\\
\ket{\psi(9)} &=\sqrt{2}\beta^{'} \ket{1_p,1_c}+\sqrt{2}\alpha^{'} \ket{3_p,0_c} =\ket{\psi(1)} \longrightarrow \text{Partially entangled} (E_{av}=0.557),
\\
\ket{\psi(10)} &=\alpha^{'} \ket{2_p,0_c}-\beta^{'}\ket{2_p,1_c} +\alpha^{'}\ket{0_p,1_c} +\beta^{'} \ket{0_p,0_c} \longrightarrow\text{MESPS } (E_{av}=1),
\end{split}
\end{align}

where, $\alpha^{'}=\frac{1}{2}(\cos(\frac{\theta}{2})-\sin(\frac{\theta}{2}))$ and $\beta^{'}=\frac{1}{2}(\cos(\frac{\theta}{2})+\sin(\frac{\theta}{2}))$. Clearly, $\ket{\psi(2)}$, $\ket{\psi(6)}$ and $\ket{\psi(10)}=\ket{\psi(2)}$ are MESPS, and we see that the sequence $H_{4}H_{4}H_{4}...$ yields MESPS with period 4, as shown in Fig.~3 of the main text.

\begin{figure}[h]
\centering
\includegraphics[width=1\linewidth]{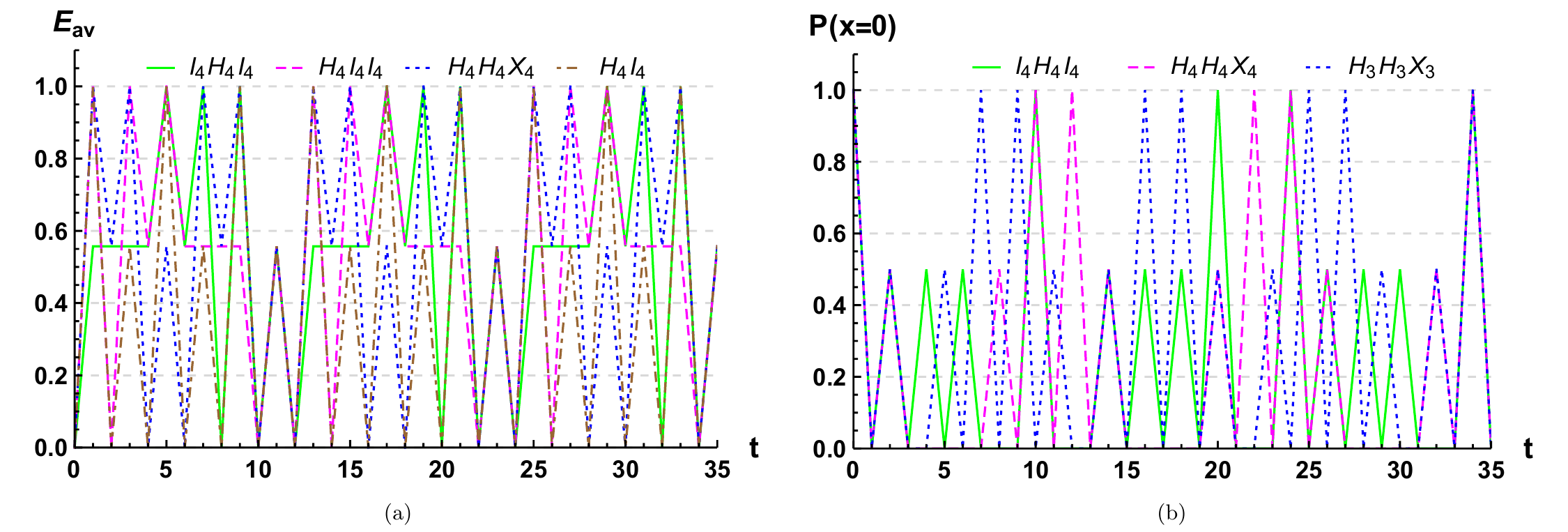}
\caption{(a) $E_{av}$ versus time steps($t$) with sequences: $I_{4}H_{4}I_{4}...$ (solid green), $H_{4}I_{4}I_{4}...$ (dashed magenta), $H_{4}I_{4}...$ (dot-dashed brown), $H_{4}H_{4}X_{4}...$ (dotted blue), $H_{4}X_{4}...$(coincides with $H_{4}I_{4}...$), for 4-cycle. (b) Probability $P(x=0)$ of finding the walker at position $\ket{0_p}$ as function of time steps($t$) with the evolution sequences $I_{4}H_{4}I_{4}...$ (solid green), $H_{4}H_{4}X_{4}...$ (dashed magenta) for 4-cycle, and $H_{3}H_{3}X_{3}...$ (dotted blue) for 3-cycle (with $\theta=0$). For both (a) and (b), $\phi=\frac{\pi}{2}$ is used.}
\label{sf2}
\end{figure}

\subsubsection{Effective-single coin and two-coin evolution sequences}
Fig.~\ref{sf2}(a) shows that the effective-single coin evolution sequences $I_{4}H_{4}I_{4}...$, $H_{4}I_{4}I_{4}...$ and $H_{4}I_{4}H_{4}I_{4}...$ yield periodic MESPS with periods 12, 12 and 4 respectively, from the separable initial state, i.e., $\ket{\psi(t=0)} = \cos(\frac{\theta}{2})\ket{0_p,0_c} + i\sin(\frac{\theta}{2})\ket{0_p,1_c}$, i.e., Eq.~(1) of the main manuscript with $\phi=\frac{\pi}{2}$. An analytical proof for the ordered QW generated by sequence $I_{4}H_{4}I_{4}...$, see Fig.~\ref{sf2}(b), is provided in the main text, and this QW periodicity supports the recurrent MESPS generation by the sequence. Fig.~\ref{sf2}(a) also shows the periodic MESPS generated by the two-coin evolution sequences $H_{4}H_{4}X_{4}...$ and $H_{4}X_{4}H_{4}X_{4}...$ (from the same initial state).

We also observe that the two-coin evolution sequence $H_{4}H_{4}X_{4}...$ gives recurring MESPS periodic in time for the 4-cycle, see Fig.~\ref{sf2}(a). This periodicity is supported by ordered QW dynamics of $H_{4}H_{4}X_{4}...$ which is shown in Fig.~\ref{sf2}(b). The occurrence of periodicity in quantum-walk of an effective single-coin or two-coin evolution sequence can be analytically proved via that for an equivalent single-coin evolution sequence. Therein, by equating the block-eigenvalues of two-coin (or, effective-single coin) evolution with those of the equivalent single-coin evolution, we find the parameters $\{\rho,\eta,\gamma\}$ of the single-coin which yields periodic QW. Here, for the  $H_{4}H_{4}X_{4}...$ sequence, the  analytical proof  following the steps involved in the periodicity condition, begins with the eigenvalues of the $U_{4,1}$ block of the evolution operator $(U_4)^3$,
\begin{equation}
\begin{split}
\lambda^{U_{4}U_{4}U_{4}}_{4,1}=\frac{1}{2}i\sqrt{\rho}e^{\frac{3}{2}i(\gamma+\eta)}(e^{-\frac{1}{2}i(\gamma+\eta)}+e^{\frac{1}{2}i(\gamma+\eta)})
(-3+2\rho+(e^{-i(\eta+\gamma)}+e^{i(\gamma+\eta)})\rho),
\label{eq10}
\end{split}
\end{equation}

then for sequence $H_{4}H_{4}X_{4}$, we have,

\begin{equation}
\lambda^{H_{4}H_{4}X_{4}}_{4,1}=\frac{\lambda^{+}_{4,1}+\lambda^{-}_{4,1}}{2}= -i,
\label{eq11}
\end{equation}
from which we get for $\delta=(\gamma+\eta)=0$, $\rho=\frac{1}{4}$, which is an exact match to the value noted in Ref.~\cite{cb-14} to generate an ordered QW on a 4-cycle (with periodicity $N=12$). Note that in Eq.~(\ref{eq11}), the used block-eigenvalues are $\lambda^{\pm}_{4,1}=-i$ and clearly, for N=12 (i.e., the period for the QW via the evolution sequence $H_4H_4X_4H_4H_4X_4...$ with $3$ steps ($v$) as the evolution $H_4H_4X_4$ repeats after every 3-time steps), we get $(\lambda^{\pm}_{4,1})^{\frac{N}{v}}=(\lambda^{\pm}_{4,1})^4=1$. In other words, the period of the QW via the evolution sequence $H_4H_4X_4...$ is, $N=4\text{(the block-eigenvalue exponent)}\times3 \text{(the number of steps $v$)}=12$. Thus, $H_{4}H_{4}X_{4}...$ renders periodic QW as shown in Fig.~\ref{sf2}(b).

\begin{figure}[H]
\centering
\includegraphics[width=1\linewidth]{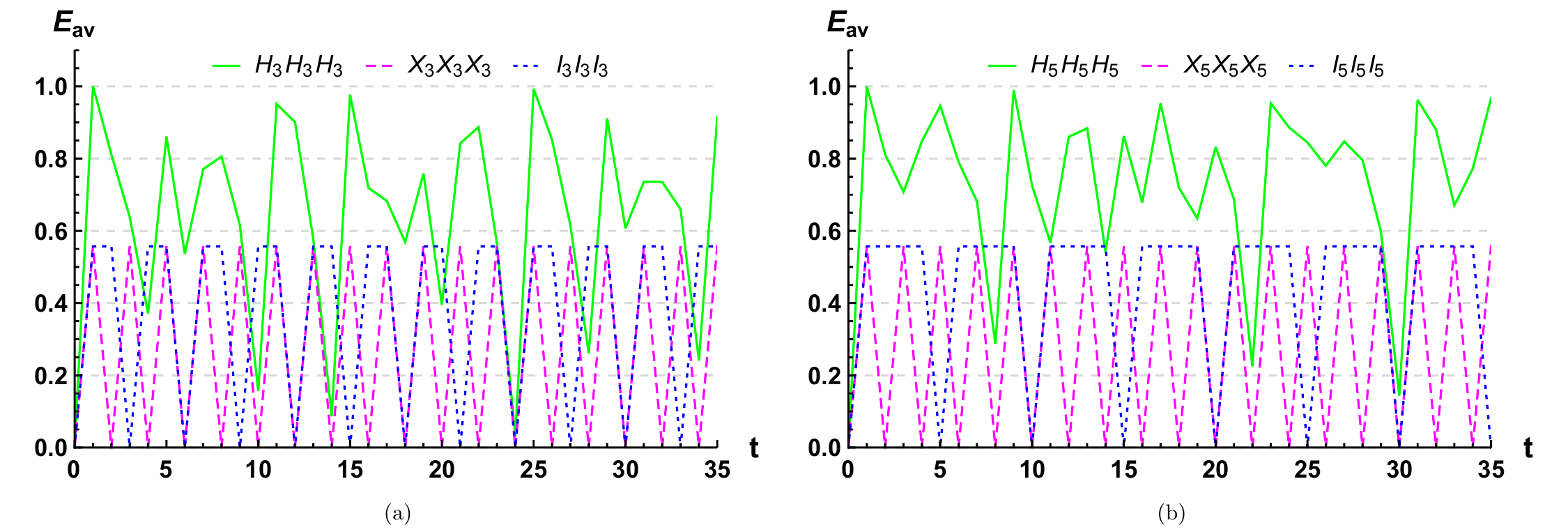}
\caption{ $E_{av}$ as function of time steps($t$) with single coin evolution sequences: (a) $H_{3}H_{3}H_{3}...$ (solid green), $X_{3}X_{3}X_{3}...$ (dashed magenta), $I_{3}I_{3}I_{3}...$ (dotted blue), for 3-cycle; (b) $H_{5}H_{5}H_{5}...$ (solid green), $X_{5}X_{5}X_{5}...$ (dashed magenta), $I_{5}I_{5}I_{5}...$ (dotted blue), for 5-cycle, with $\phi=\frac{\pi}{2}$.}
\label{sf3}
\end{figure}

\subsection{\large 3-cycle and 5-cycle}
\subsubsection{Single-coin evolution sequences}
Results for DTQW on 3- and 5-cycles yield effective single-coin evolution sequences and two-coin evolution sequences that result in periodic MESPS, but no single-coin evolution sequence was found for the $k\in\{3,5\}$-cycles. However, sequence $H_{k}H_{k}H_{k}...$ gives MESPS only at time step $t=1$ for $k=3$ and $k=5$-cycles, see Fig.~\ref{sf3} (for the separable initial state with $\phi=\frac{\pi}{2}$). This is unlike the sequence $H_{4}H_{4}H_{4}...$ for the 4-cycle case, which gives MESPS at $t=1,5,9...$ with period $4$, as discussed in the main text, see also Fig.~\ref{sf1}(a).

\subsubsection{Effective-single coin and two-coin evolution sequences}
\begin{figure}[H]
\centering
\includegraphics[width=1\linewidth]{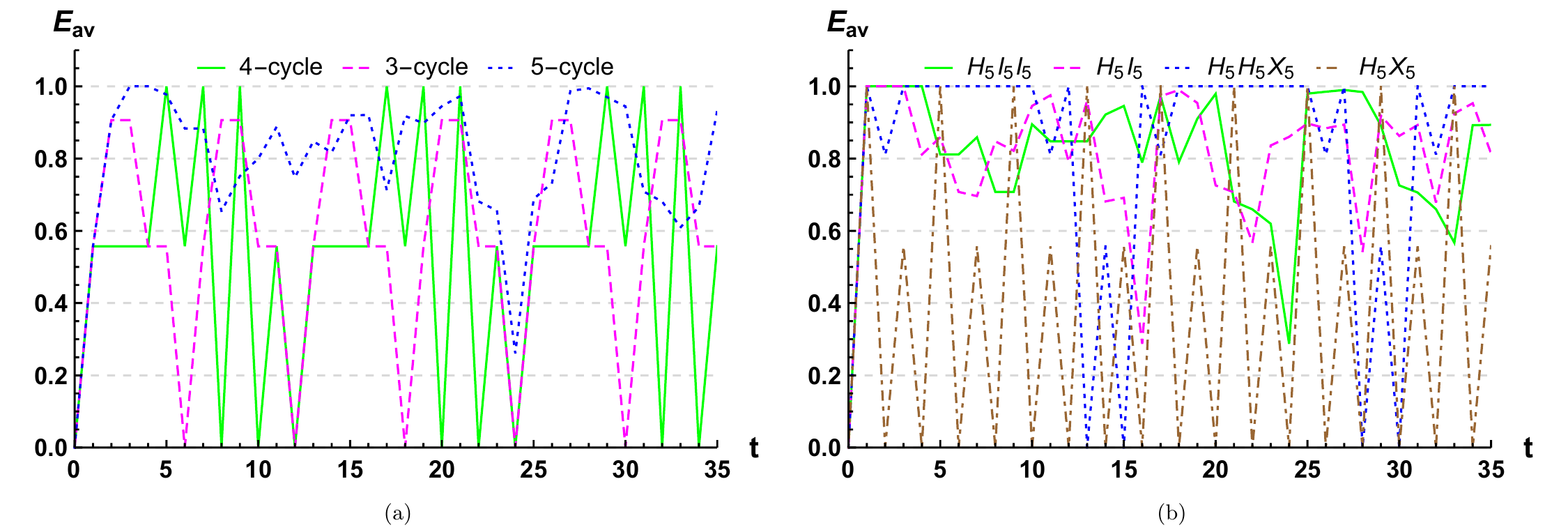}
\caption{(a) $E_{av}$ for effective-single coin evolution sequence $I_{k}H_{k}I_{k}...$ up to 35 time steps($t$), for $k\in\{3,4,5\}$ i.e., 3-cycle (dashed magenta), 4-cycle (solid green) and 5-cycle (dotted blue). (b) $E_{av}$ versus time steps($t$) with sequences: $H_{5}I_{5}I_{5}...$ (solid green), $H_{5}I_{5}...$ (dashed magenta), $H_{5}H_{5}X_{5}...$ (dotted blue), $H_{5}X_{5}...$ (dot-dashed brown), for 5-cycle. For both (a) and (b), $\phi=\frac{\pi}{2}$ is used.}
\label{sf4}
\end{figure}

Fig.~\ref{sf4}(a) shows that for the separable initial state Eq.~(\ref{eq1}) with $\phi=\frac{\pi}{2}$, the effective single coin sequence $I_{3}H_{3}I_{3}...$ on 3-cycle renders ordered QW without MESPS, whereas sequence $I_{5}H_{5}I_{5}...$ on 5-cycle renders chaotic QW but with MESPS at $t=3,4$. This is unlike the sequence $I_{4}H_{4}I_{4}...$ on 4-cycle, which generates recurrent MESPS with period 12.

Fig.~\ref{sf4}(b) shows that with the same initial state on 5-cycle, effective-single coin sequences $H_{5}I_{5}I_{5}..., $ and $H_{5}I_{5}H_{5}I_{5}...$ do not give periodic MESPS, but yield MESPS at $t=1,2,3,4$ and $t=1,2,3$ respectively.

We also observe that the two-coin evolution sequences $H_{k}H_{k}X_{k}...$ and $H_{k}X_{k}H_{k}X_{k}...$ yield recurring and periodic MESPS for both $k=3$- and $k=5$-cycles for the separable initial state with $\phi=\frac{\pi}{2}$, see Fig.~\ref{sf4}(b) and main text Fig.~5.

A similar analytical proof for the periodic behavior of $H_{3}H_{3}X_{3}...$ in generating recurrent MESPS on the 3-cycle can also be shown. Firstly, we get for the eigenvalues of the $U_{3,1}$ block of the evolution operator $(U_3)^3$,
\begin{equation}
\begin{split}
\lambda^{U_{3}U_{3}U_{3}}_{3,1}=\frac{1}{4}\sqrt{\rho}(3(1+i\sqrt{3})e^{i(\eta+\gamma)}(\rho-1)+3i(i+\sqrt{3})e^{2i(\eta+\gamma)}(\rho-1)
+2\rho-2e^{3i(\eta+\gamma)}\rho),
\label{eq12}
\end{split}
\end{equation}

and then for the sequence $H_{3}H_{3}X_{3}$, we have,
\begin{equation}
\lambda^{H_{3}H_{3}X_{3}}_{3,1}=\frac{\lambda^{+}_{3,1}+\lambda^{-}_{3,1}}{2}=-\frac{i \sqrt{3}}{2},
\label{eq13}
\end{equation}
from which
$\rho= 0.550901,\; 0.15597\;$ with $(\gamma+\eta)=0$, which match the values noted in Ref.~\cite{cb-14} to generate an ordered QW on a 3-cycle (with periodicity $N=18$). 
Also, the block-eigenvalues used in Eq.~(\ref{eq13}) are $\lambda^{+}_{3,1}=-\frac{1}{2}(1+i\sqrt{3})$, $\lambda^{-}_{3,1}=\frac{1}{2}(1- i\sqrt{3})$ and for $N=18$ , we get $(\lambda^{\pm}_{3,1})^{\frac{N}{v}}=(\lambda^{\pm}_{3,1})^{\frac{18}{3}}=(-\frac{1}{2}(1+i\sqrt{3}))^6=(\frac{1}{2}(1-i\sqrt{3}))^6=1$ (here, $v=3$ is the number of steps in the evolution $H_{3}H_{3}X_{3}$ which repeats itself after every 3-time steps in the evolution sequence $H_{3}H_{3}X_{3}H_{3}H_{3}X_{3}...$).
Thus, the sequence $H_{3}H_{3}X_{3}...$ renders a periodic QW  dynamics with period 18, as shown in Fig.~\ref{sf2}(b). See Fig.~5 of the main text for recurrent MESPS generated via $H_{3}H_{3}X_{3}...$.

Furthermore, one can see from Fig.~\ref{sf4}(b) that in the 5-cycle case, the sequences $H_{5}H_{5}X_{5}...$ and $H_{5}X_{5}H_{5}X_{5}...$ yield MESPS respectively at $t=1,3,4,5,6,7,8,9,10,12,16,...$ with period 15 and at $t=1,5,9,...$ with period 4. Moreover, it is interesting to observe that with just sequences $H_{5}H_{5}X_{5}...$ and $H_{5}I_{5}I_{5}...$, one can generate MESPS for all time steps $t\le10$ and also at larger $t$, on a 5-cycle.

\newpage
\section{Evolution sequences to generate MESPS up to $10$-time steps and beyond with cyclic graphs}
\label{App.C}
The results on MESPS generation from the separable initial state Eq.~(\ref{eq1}) with $\phi\in\{\frac{\pi}{2},\frac{\pi}{6},\pi\}$, are juxtaposed in Table~\ref{table1p2}, where one can compare the proposed evolution sequences on cyclic graphs to generate recurring or periodic MESPS at time steps up to 10 and beyond. We see that with $\phi=\frac{\pi}{2}$, employing $H_{3}H_{3}X_{3}...,\; H_{3}I_{3}I_{3}..., $ and $H_{3}X_{3}H_{3}X_{3}...$ on a 3-cycle one can obtain MESPS at all time steps up to 10, whereas on a 4-cycle their analogs give MESPS at all odd time steps $t\le10$, see Fig.~\ref{sf2}(a) and Fig.~5 of the main text. As these sequences also beget periodic QWs. Thus, one obtains MESPS at larger time steps ($t>10$) too. Moreover, it is interesting to observe that with the sequences $H_{5}H_{5}X_{5}...$ and $H_{5}I_{5}I_{5}...$, one can generate MESPS at all time steps $t\le10$ and also at larger $t$, on a 5-cycle, see Fig.~\ref{sf4}(b). In fact, $H_{5}H_{5}X_{5}...$ by itself yields MESPS at time steps $t=1,3,4,5,6,7,8,9,10,12,16,...$ with period 15.

\begin{table*}[h!]
\caption{ \normalsize Evolution sequences to generate MESPS via DTQW up to $10$-time steps and beyond periodically with $k\in\{3,4,5,8\}$-cycles}
\begin{tabular}{ |c|l|l| }
\hline
\textbf{Initial state ($\phi$ value)} & \textbf{MESPS on 4-cycle and 8-cycle} & \textbf{MESPS on 3-cycle and 5-cycle} \\
\hline
\multirow{14}{4em}{$\phi=\frac{\pi}{2}$ } & \underline{Single coin evolution sequences:} & \underline{Single coin evolution sequences:} \\
& $H_{4}H_{4}H_{4}..,C^{'}_{4}C^{'}_{4}C^{'}_{4}..$ at $t=1,5,9,..$ ($P=4$) & $H_{3}H_{3}H_{3}...$ at $t=1$ (Chaotic)\\
& $F_{4}F_{4}F_{4}...$ at $t=2,6,10,...$ ($P=4$) &$H_{5}H_{5}H_{5}...$ at $t=1$ (Chaotic) \\
& $C^{'}_{4}C^{'}_{4}C^{'}_{4}...$ at $t=1,5,9,...$ ($P=4$) & \underline{Effective single coin evolution sequences:}\\
& $C_{4}C_{4}C_{4}...$ at $t=5,17,29,...$ ($P=12$) & $H_{3}I_{3}I_{3}...$ at $t=1,2,7,8,...$ $(P=6)$ \\
& $H_{8}H_{8}H_{8}..,C^{'}_{8}C^{'}_{8}C^{'}_{8}..$ at $t=1,13,25,...$ ($P=12$)& $H_{3}I_{3}H_{3}I_{3}...$ at $t=1,2$ (Chaotic)\\
& \underline{Effective single coin evolution sequences:} & $I_{5}H_{5}I_{5}...$ at $t=3,4$ (Chaotic) \\
& $I_{4}H_{4}I_{4}...$ at $t=5,7,9,17,...$ ($P=12$)&$H_{5}I_{5}I_{5}...$ at $t=1,2,3,4$ (Chaotic) \\
& $H_{4}I_{4}I_{4}...$ at $t=1, 3, 5, 13,...$ $(P=12)$ & $H_{5}I_{5}H_{5}I_{5}...$ at $t=1,2,3$ (Chaotic)\\
& $H_{4}I_{4}H_{4}I_{4}...$ at $t=1, 5, 9,...$ $(P=4)$ & \underline{Two coin evolution sequences:}\\
& \underline{Two coin evolution sequences: }& $H_{3}H_{3}X_{3}...$ at $t=1,3,4,6,10,...$ $(P=9)$\\
& $H_{4}H_{4}X_{4}...$ at $t=1,3,7,9,13,...$ $(P=6)$&$H_{3}X_{3}H_{3}X_{3}...$ at $t=1,5,9,...$ $(P=4)$ \\
&
$H_{4}X_{4}H_{4}X_{4}...$ at $t=1,5,9,...$ $(P=4)$ & $H_{5}H_{5}X_{5}...$ at $t=1,3-10,12,16,...$ $(P=15)$\\
& &$H_{5}X_{5}H_{5}X_{5}...$ at $t=1,5,9,...$ $(P=4)$ \\

\hline
\multirow{4}{4em}{$\phi=\frac{\pi}{6}$}
& \underline{Single coin evolution sequences:} &\underline{Single coin evolution sequences:}\\
& $R_{4}R_{4}R_{4}...$ at $t=1,5,9,...$ ($P=4$) & $R_{3}R_{3}R_{3}...$ at $t=1$ (Chaotic) \\
& $Q_4Q_4Q_4...$ at $t=1,4,7...$ ($P=3$) & $Q_3Q_3Q_3..., Q_5Q_5Q_5...$ at $t=1$ (Chaotic) \\
& $R_{8}R_{8}R_{8}...$ at $t=1,13,25,...$ ($P=12$)& $R_5R_5R_5...$ at $t=1$ (Chaotic) \\
\hline
\multirow{4}{4em}{$\phi=\pi$} &\underline{Single coin evolution sequences:} &\underline{Single coin evolution sequences:}\\
& $F_{4}F_{4}F_{4}...$ at $t=1,5,9,...$ ($P=4$) & $F_3F_3F_3...$ at $t=1$ (Chaotic) \\
& $H_{4}H_{4}H_{4}...$ at $t=2,6,10,...$ ($P=4$) & $F_5F_5F_5...$ at $t=1$ (Chaotic) \\
& $F_{8}F_{8}F_{8}...$ at $t=1,13,25,...$ ($P=12$) & \\
\hline
\multicolumn{3}{|c|}{$t\rightarrow$ time steps, $P\rightarrow$ Period at which the sequence yields maximal SPE (MESPS)}\\
\hline
\end{tabular}
\label{table1p2}
\end{table*}

\section{Security of the MESPS based cryptographic protocol}
\label{App.D}

We exploit our MESPS generation scheme via periodic DTQW for a cryptographic protocol, wherein MESPS is a public key for encoding a secret message ($m$), and the decryption requires a measurement based on the evolution operator sequence generating the periodic MESPS. Ref.~\cite{crypt15} shows a simpler cryptographic protocol which uses inverse evolution operation for message decryption.

As described in the main letter, Alice sends the message
$m\in{\{0,1,2,3\}}$ to Bob by exploiting the MESPS generation via single-coin evolution sequence $H_4H_4H_4...$
on a cyclic graph with $4$ sites. This protocol involves the following three steps.--

Step-1: Bob generates the public key as $\ket{\psi_{pk}}=A\ket{j_p}\ket{q_c}$, where $A=(H_{4})^5$ and, $\ket{j_p}$ with $j\in\{0,1,2,3\}$ and $\ket{q_c}=\cos(\frac{\theta}{2})\ket{0_c}+i\sin(\frac{\theta}{2})\ket{1_c}$ are respectively the position and coin states of the quantum walker. $\{A,j\}$ is the secret or private key~\cite{crypt15}. After generating this MESPS $|\psi_{pk}\rangle$ that acts as the public key, Bob sends it to Alice.

Step-2: Alice then encodes the message via: $\ket{\psi(m)}=(T_m\otimes I_c)\ket{\psi_{pk}}, \text{where } T_m=\sum^{3}_{i=0}\ket{((i+m)\text{ mod } 4)_p}\bra{i_p},$ acts on position state and $I_c$ is the $2\times2$ identity operator acting on coin state, and sends it to Bob.

Step-3: Bob now decrypts the message by operating $W=(H_{4})^{3}$ from which he gets $\ket{((j+m)\text{ mod }4)_p,q_c}$. Bob reads $m{'}=(j+m)\text{ mod}\;4$, from the position ket and from which he obtains Alice's message $m$.
\noindent

On the security of our proposed cryptographic protocol, let us consider that the position state be $\ket{j_p}=\ket{0_p}$, and coin state $\ket{q_c}=\ket{0_c}$ initially and the message is, say $m=3$.

The public key from step-1 now is $\ket{\psi_{pk}}=A \ket{0_p,0_c}=\frac{1}{\sqrt{2}}(\ket{1_p,0_c}+\ket{3_p,1_c})$, with $A=(H_{4})^5$, which is a MESPS (Bell state), see the state at $t=5$ of Eq.~\eqref{eqn:eq7}. An eavesdropper can attack at step-2 when Alice sends the encrypted and encoded message $\ket{\psi(m)}$ to Bob.

The message is encoded as in step-2, by Alice and is given by $\ket{\psi(m=3)}=(T_3\otimes I_c)\ket{\psi_{pk}}=\frac{1}{\sqrt{2}}[\ket{((1+3)\text{ mod }4)_p,0_c}+\ket{((3+3)\text{ mod }4)_p,1_c}]=\frac{1}{\sqrt{2}}[\ket{0_p,0_c}+\ket{2_p,1_c}]$.

In the absence of the eavesdropping, Bob will decrypt the message from $\ket{\psi(m=3)}$ by operating $W=(H_{4})^{3}$. On operating $W$, Bob obtains $\ket{((0+3)\text{ mod }4)_p,0_c}=\ket{3_p,0_c}$ and from which Bob gets $m{'}=3\text{ mod }4=3$, and from which he securely obtains Alice's message $m=3$.

Now, let us suppose an eavesdropper (Eve) is present at step-2. Since Eve does not know the private key i.e., $\{A,j\}$, Eve learning the state $\ket{\psi(m=3)}$ is almost impossible, (i.e., it has negligible probability~\cite{crypt15}). This mitigates an intercept-and-resend attack~\cite{qkd-attack2020} wherein Eve can try to measure Alice’s signal or the message-encrypted state $\ket{\psi(m=3)}$. It is noteworthy that robust maximal entanglement via MESPS in the public key $\ket{\psi_{pk}}$ as compared to a superposed or product state (as public key), makes it even harder for an eavesdropper to extract information from the public key by means of POVM measurements~\cite{crypt15}. Further, the challenge in the receiver’s authentication (i.e., a man-in-the-middle attack~\cite{qkd-attack2020}), i.e., whether the receiver is a friend (Bob) or a foe like Eve, is taken care of by the pre-shared private key $\{A, j\}$ among Alice and Bob. Additionally, Eve does not know which operator is required to apply on $\ket{\psi(m)}$ to retrieve the message sent by Alice. Further, Eve's probability of guessing $W$ and retrieving back the state $\ket{\psi(m=3)}$ is negligible, as there exist an infinite number of possibilities for the coin and its combinations. 

To conclude, we note that the private key is unknown to Eve, and Eve has a negligible probability of learning Alice's prepared state $\ket{\psi(m=3)}$. Additionally, seeing the infinite possibilities for $2\times2$ coins and their combinations for generating an evolution operator, Eve guessing exact $W$ and getting back the state $\ket{\psi(m=3)}$, is impossible. Therefore, our proposed cryptographic protocol is secure and foolproof against any attack. Though our protocol is secure, adopting techniques like hardware adaptation for quantum system isolation and privacy amplification will further minimize the deviation from theory, as far as practical implementation of the cryptography scheme is concerned~\cite{qkd-attack2020}. 

This cryptography protocol can also be achieved using any evolution sequence yielding periodic MESPS (e.g., see Table~\ref{table1p2}), following the above-mentioned series of steps with any $k\in\{3,4,5,8\}$-cycle.



\section{A comparison of our scheme and results with other relevant proposals}
\label{App.E}
We compare our scheme and results with other relevant proposals~\cite{r_zhang2022, fang,me-cb,gratsea2020universal,gratsea_lewenstein_dauphin_2020} in Table~\ref{table2p2}.
We first compare the type of coin (evolution) sequences and the number of coin operators used. In our paper, we use a single coin, an effective-single coin, and two coin evolution sequences. Ref.~\cite{me-cb} uses deterministic Parrondo type two-coin evolution sequences and has considered four coin operators. In Refs.~\cite{gratsea2020universal,gratsea_lewenstein_dauphin_2020}, coin sequences based on optimization and a deterministic sequence, namely- universal entangler~\cite{gratsea2020universal} are proposed. Ref.~\cite{fang} uses a rigorous optimization scheme to obtain effective single coin sequences (with $\hat{H}$ and $\hat{I}$ coins) with quantum process fidelity as the cost function. Ref.~\cite{r_zhang2022} deals with inhomogeneous QW, wherein position-dependent coin operations are used. Experimental realization is straightforward with less number of coin operators being involved and with a small number of sites. Given this, our proposed sequences (single or effective-single coin evolution sequences) are as good as the entangling sequences of~\cite{gratsea2020universal,gratsea_lewenstein_dauphin_2020, fang,r_zhang2022, me-cb} and are much simpler to work with. Further, the MESPS generating schemes in Refs.~\cite{r_zhang2022,me-cb, gratsea2020universal, gratsea_lewenstein_dauphin_2020} are not independent of initial state parameters $\{\phi,\theta\}$ unlike the scheme of Ref.~\cite{fang}, whereas, our scheme is independent of the initial state parameters subject to the condition $(\gamma+\phi) \in \{\frac{\pi}{2},\frac{3\pi}{2}\}$, where $\gamma$ is a coin parameter (see Eq.~(\ref{eq6})). Moreover, our work involves only $3, 4,$ or $5$ sites for the QW evolution, which is also resource-saving. We are the first to propose single-coin evolution sequences that generate recurrent MESPS, which will be easiest to work with experimentally.

\begin{sidewaystable}
\centering
\caption{Comparison of the present work with other relevant works}
\Large
\resizebox{1\textwidth}{!}{%
\begin{tabular}{|c|l|l|l|l|l|l|l|}
\hline
\textbf{Properties$\downarrow$/Model$\rightarrow$} &
\multicolumn{1}{c|}{\textbf{\begin{tabular}[c]{@{}c@{}} This Paper\\(4-cycle with single coin:\\ $\hat{C}_{2}(\rho,\gamma,\eta) \text{ as in Eq.}~(\ref{eq6})$)
\end{tabular}}} &
\multicolumn{1}{c|}{\textbf{\begin{tabular}[c]{@{}c@{}} This Paper\\(3,4,5-cycles with effective-single coin \\or two-coin evolution sequences) \end{tabular}}} &
\multicolumn{1}{c|}{\textbf{\begin{tabular}[c]{@{}c@{}} With Parrondo \\sequences\\ Refs.~\cite{me-cb}\end{tabular}}} &
\multicolumn{1}{c|}{\textbf{\begin{tabular}[c]{@{}c@{}}With \\Optimization\\ Ref.~\cite{fang}\end{tabular}}} &
\multicolumn{1}{c|}{\textbf{\begin{tabular}[c]{@{}c@{}}Analysis with\\ inhomogeneous-QW\\ Ref.~\cite{r_zhang2022}\end{tabular}}} &
\multicolumn{1}{c|}{\textbf{\begin{tabular}[c]{@{}c@{}}With \\Optimization\\ Ref.~\cite{gratsea_lewenstein_dauphin_2020}\end{tabular}}} &
\multicolumn{1}{c|}{\textbf{\begin{tabular}[c]{@{}c@{}}With \\Optimization\\Ref.~\cite{gratsea2020universal}\end{tabular}}} \\

\hline
\hline
\textbf{\begin{tabular}[c]{@{}c@{}}No. of coin\\ operators used\end{tabular}} &
\begin{tabular}[c]{@{}l@{}} Single coin as in Eq.~(\ref{eq6}) (say, $\hat{G}$)\end{tabular} &
\begin{tabular}[c]{@{}l@{}} Effective-single coin or, 2 coins\end{tabular} &
\begin{tabular}[c]{@{}l@{}}Two-coin sequences\end{tabular} &
\begin{tabular}[c]{@{}l@{}} Effective single coin\\ (Hadamard with Identity \\coin) \end{tabular} &
\begin{tabular}[c]{@{}l@{}}2 coins \\inhomogeneously \end{tabular} &
\begin{tabular}[c]{@{}l@{}}Full set of possible\\ coin operators\end{tabular} &
\begin{tabular}[c]{@{}l@{}}2 coins\end{tabular} \\ \hline
\textbf{\begin{tabular}[c]{@{}c@{}}Procedure\end{tabular}} &
\begin{tabular}[c]{@{}l@{}}Simple,\\ single coin QW on cyclic graph\end{tabular} &
\begin{tabular}[c]{@{}l@{}}Simple, QW with deterministic\\ evolution sequences on cyclic graph \end{tabular} &
\begin{tabular}[c]{@{}l@{}}QW on 1D line with\\ Parrondo sequences\end{tabular} &
\begin{tabular}[c]{@{}l@{}}Optimization with\\ QW on 1D line\end{tabular} &
\begin{tabular}[c]{@{}l@{}}Inhomogenous QW \\on 1D line \end{tabular} &
\begin{tabular}[c]{@{}l@{}}Basin hopping algorithm,\\ QW on 1D line\end{tabular} &
\begin{tabular}[c]{@{}l@{}}RL technique,\\ QW on 1D line\end{tabular} \\ \hline

\textbf{\begin{tabular}[c]{@{}c@{}}Independent of initial\\ state parameters ($\phi,\theta$)\end{tabular}} &
\begin{tabular}[c]{@{}l@{}}Yes, subject to $\gamma+\phi\in\{\frac{\pi}{2},\frac{3\pi}{2}\}$\\ where, $\gamma$ is a coin parameter\end{tabular} &
\begin{tabular}[c]{@{}l@{}} Partially\end{tabular} &
\begin{tabular}[c]{@{}l@{}}Partially\end{tabular} &
\begin{tabular}[c]{@{}l@{}}Yes\end{tabular} &
\begin{tabular}[c]{@{}l@{}} No\end{tabular} &
\begin{tabular}[c]{@{}l@{}}No\end{tabular} &
\begin{tabular}[c]{@{}l@{}}Partially\end{tabular} \\ \hline

\textbf{\begin{tabular}[c]{@{}c@{}}Maximally \\entangled states\end{tabular}} &
\begin{tabular}[c]{@{}l@{}}
Infinitely many single coins ($\hat{G}$) as in Eq.~(\ref{eq6})\\ yield MESPS at $t=1$, on any $k-cycle$ .\\
$\hat{G}$ with $(\gamma+\eta)\in\{0,\pi,\frac{\pi}{2},\frac{3\pi}{2}\}$ yields \\recurrent MESPS on 4- and 8-cycles.\\
Example: At time steps,
\\
$t=1,5,9,...(\text{with } H_{4}H_{4}H_{4}...,\;P=4)$.
\\
Moreover at,\\$t=5,17,29,...$(\text{with } $C_{4}C_{4}C_{4}...,\;P=12$).\\
($P\rightarrow$ Period at which the evolution\\ sequence yields MESPS.)\\

\end{tabular} &
\begin{tabular}[c]{@{}l@{}}At time steps($t$), \\
$t=$1,3,4,6,10,... ($H_{3}H_{3}X_{3}...,\;P=9$);\\
$t=$1,2,7,8... ($H_{3}I_{3}I_{3}...,\;P=6$);\\
$t=1,5,9,...$ ($H_{k}X_{k}H_{k}X_{k}...,\;k=3,4,5$,\;$P=4$);
\\
moreover, at\\ $t=$1,3-10,12,16,... ($H_{5}H_{5}X_{5}...,\;P=15$);\\
$t=$1,2,3,4 ($H_{5}I_{5}I_{5}...$,\;Chaotic);\\
$t=$5,7,9,17,... ($I_{4}H_{4}I_{4}...$,\;$P=12$), etc.
\\So MESPS $\forall\;t\le10$ and larger $t$. \\And periodic emergence of MESPS.\\

\end{tabular} &
\begin{tabular}[c]{@{}l@{}}At $t=3,5$\\ and at asymptotic $t$ \end{tabular} &
\begin{tabular}[c]{@{}l@{}}At $t=3$\\ and beyond \end{tabular} &
\begin{tabular}[c]{@{}l@{}}At any odd $t$ and \\asymptotically in even t \end{tabular} &
\begin{tabular}[c]{@{}l@{}}Almost at $t=10$\\ and beyond \end{tabular} &
\begin{tabular}[c]{@{}l@{}}Not achieved \end{tabular}\\
\hline
\end{tabular}%
}
\label{table2p2}
\end{sidewaystable}

Additionally, focusing on a small number of time steps in Ref.~\cite{gratsea_lewenstein_dauphin_2020} shows maximal entanglement can be achieved in $10$-time steps and beyond. Ref.~\cite{fang} generates maximal entanglement via the optimization problem for any time step beyond the second, whereas Ref.~\cite{r_zhang2022} shows maximal entanglement can be generated for any odd time steps, and in the asymptotic limit for even steps. The method proposed in Ref.~\cite{me-cb} gives MESPS in $3$ and $5$ time steps independent of the initial states. For the first time, we achieve
a framework for arbitrary single coins which yield recurrent MESPS (starting from time step $t=1$). In addition, Our scheme shows the generation of MESPS at all time steps ($t$) up to $10$ and at a larger $t$ with periodic occurrence.

\newpage
\onecolumngrid
\section{Python code}
\label{App.F}
Herein, we provide a typical Python code that can generate Fig. 4 ($H_4H_4H_4...$ sequence, shown in solid green) of the main text (or any figure of the manuscript) for curious and interested readers.

\widetext
{\onecolumngrid
\begin{scriptsize}
\begin{verbatim}
PYTHON CODE:

from numpy import *
import numpy as np
import math
import random
from scipy import integrate
import matplotlib.pyplot as mp
N = 100
DKP = np.zeros(N)
pi=np.pi
cos=np.cos
sin=np.sin
sqrt=np.sqrt
eye=np.eye
roll=np.roll
kron=np.kron
zeros=np.zeros
exp=np.exp
empty=np.empty
outer=np.outer
phi=pi/2
coin0 = np.array([1, 0])
coin1 = np.array([0, 1])
C00 = np.outer(coin0, coin0)
C01 = np.outer(coin0, coin1)
C10 = np.outer(coin1, coin0)
C11 = np.outer(coin1, coin1)
C_hat = (cos(pi/4)*C00 + sin(pi/4)*C01 + sin(pi/4)*C10 - cos(pi/4)*C11)
def Ent(l, phi,P,ShiftPlus,ShiftMinus,S_hat,H):
posn0 = zeros(P)
posn0[0] = 1
psi0 = kron(posn0,(cos(l/2)*coin0 + exp(1j*phi)*sin(l/2)*coin1))
psiN = psi0 #Initialisation
for i in range(1, m + 1):
psiN = np.linalg.matrix_power(H,1).dot(psiN)
prob1 = empty(P)
prob2 = empty(P)
prob3 = empty(P)
prob = empty(P)
for k in range(P):
posn = zeros(P)
posn[k] = 1
M_hat_k = kron(outer(posn,posn), eye(2))
X_hat_k = kron(outer(posn,posn), C00)
W_hat_k = kron(outer(posn,posn), C11)
proj = M_hat_k.dot(psiN)
proj1 = X_hat_k.dot(psiN)
proj01 = X_hat_k.dot(kron(posn,(coin0)))
proj2 = W_hat_k.dot(psiN)
proj02 = W_hat_k.dot(kron(posn,(coin1)))
prob1[k] = (proj02.dot(proj2.conjugate())*proj01.dot(proj1.conjugate()).conjugate()).real
prob2[k] = (proj02.dot(proj2.conjugate())*proj01.dot(proj1.conjugate()).conjugate()).imag
prob3[k] = ((proj1.dot(proj1.conjugate())).real - (proj2.dot(proj2.conjugate())).real)/2
n1 = n2 = n3 = 0
for j in range(P):
n1 = prob1[j] + n1
n2 = prob2[j] + n2
n3 = prob3[j] + n3
n = sqrt(n1**2 + n2**2 + n3**2)
Ev1 = (0.5 - round(n,14))
Ev2 = (0.5 + round(n,14))
if Ev1==0 :
VEnt = -Ev2*math.log2(Ev2)
else:
VEnt = -Ev1*math.log2(Ev1)-Ev2*math.log2(Ev2)
VEntr= VEnt/pi
return VEntr
k1=4
out1=outer(zeros(4),zeros(4))
for i1 in range(0,k1):
out1=out1+outer(eye(1,4,np.mod(i1-1,k1)),eye(1,4,i1))
out2=outer(zeros(4),zeros(4))
for i2 in range(0,k1):
out2=out2+outer(eye(1,4,np.mod(i2+1,k1)),eye(1,4,i2))
for m in range(1, N + 1):
P=4
ShiftPlus = roll(out2, 1, axis=0)
ShiftMinus = roll(out1, -1, axis=0)
S_hat = kron(out2, C11) + kron(out1, C00)
H = S_hat.dot(kron(eye(P), C_hat))
ans, err = integrate.quad(lambda l: Ent(l, phi,P,ShiftPlus,ShiftMinus,S_hat,H), 0, pi)
DKP[m-1] = ans
print(DKP[m-1]) #Print average entanglement entropy with Hadamard coin up to N time steps, for 4-cycle:

\end{verbatim}
\end{scriptsize}}

\end{document}